\documentstyle[epsf,12pt]{article}

\begin{document}

\title
{Three-loop $\beta$-function for $N=1$ supersymmetric
electrodynamics, regularized by higher derivatives.}

\author{A.A.Soloshenko \thanks{E-mail:$solosh@theor.phys.msu.su$},
K.V.Stepanyantz \thanks{E-mail:$stepan@theor.phys.msu.su$}}

\maketitle

\begin{center}

{\em Moscow State University, Physical Faculty,\\
Department of Theoretical Physics.\\$117234$, Moscow, Russia}

\end{center}

\begin{abstract}
Three-loop quantum corrections to the effective action are calculated
for $N~=~1$ supersymmetric electrodynamics, regularized by higher
derivatives. Using the obtained results we investigate the anomaly
puzzle in the considered model.
\end{abstract}

\sloppy


\section{Introduction.}
\hspace{\parindent}

It is well known \cite{Ferrara,Clark,Piquet1,Piquet2}, that in
supersymmetric theories the axial and the trace of the energy-momentum
tensor anomalies are components of a chiral scalar supermultiplet.
Adler-Bardeen theorem \cite{Bardeen,Slavnov_Book} asserts that there
are no radiative corrections to the axial anomaly beyond the one-loop
approximation, while the trace anomaly is proportional to the
$\beta$-function \cite{Adler_Collins} to all orders. Therefore it
seems to imply, that the $\beta$-function in supersymmetric theories
should be exhausted by the first loop \cite{NSVZ_PL}. It does take
place in models with $N=2$ supersymmetry \cite{N2}. However explicit
perturbative calculations find higher order corrections to the
$\beta$-functions of $N=1$ supersymmetric theories, regularized by
dimensional reduction \cite{Tarasov,Grisaru,Caswell}. This contradiction
is usually called "the anomaly puzzle".

Many papers were written in the attempt of solving the anomaly puzzle in
supersymmetric theories. For example, in \cite{SV} the anomaly puzzle
is argued to be a consequence of the difference between the usual and
Wilsonian effective actions. In particular, the authors noted, that there
was a nontrivial contribution to the $\beta$-function related with the
Konishi anomaly \cite{Konishi,ClarkKonishi}. The investigation of this
contribution in \cite{SV} and the investigation of instanton contributions
in \cite{NSVZ_Instanton} have led to the construction of the so-called
exact Novikov, Shifman, Vainshtein and Zakharov (NSVZ) $\beta$-function.
For $N=1$ supersymmetric electrodynamics (SUSY QED) considered in this
paper the NSVZ $\beta$-function has the following form:

\begin{equation}\label{NSVZ_Beta}
\beta(\alpha) = \frac{\alpha^2}{\pi}\Big(1-\gamma(\alpha)\Big),
\end{equation}

\noindent
where $\gamma(\alpha)$ is the anomalous dimension of the matter superfield.
Explicit perturbative calculations with the dimensional reduction (DRED)
verify the NVSZ $\beta$-function up to the two-loop order. Nevertheless,
the three-loop results obtained in \cite{ThreeLoop1,ThreeLoop2,ThreeLoop3}
do not agree with the NSVZ $\beta$-function. However \cite{ThreeLoop2}
this disagreement can be eliminated by a special choice of renormalization
scheme, the possibility of such a choice being highly nontrivial
\cite{JackJones}. Actually it is possible to relate
$\overline{\mbox{DRED}}$ scheme and NSVZ scheme order by order
\cite{North} in the perturbation theory. It is worth mentioning, that at
two-loops the NSVZ $\beta$-function was also obtained with differential
renormalization \cite{DiffR}. For example, for $N=1$ SUSY Yang-Mills
the calculation was made in \cite{Mas}.

However the relation between $\Gamma$ and the Wilsonian action remained
unclear. This problem was avoided in another solution of the anomaly puzzle,
proposed in \cite{Arkani}. The main idea of \cite{Arkani} is that the higher
order corrections in NSVZ $\beta$-function are due to anomalous Jacobian
under the rescaling of the fields done in passing from holomorphic to
canonical normalization. In the case of supersymmetric electrodynamics
holomorphic normalization means, that the renormalized action is written as

\begin{eqnarray}\label{Holomorphic_Normalization}
&& S_{ren} =
\frac{1}{4 e^2} Z_3(\Lambda/\mu) \mbox{Re}
\int d^4x\,d^2\theta\,W_a C^{ab} W_b
+\nonumber\\
&& \qquad\qquad\qquad\qquad\qquad
+ Z(\Lambda/\mu)\frac{1}{4}\int d^4x\, d^4\theta\,
\Big(\phi^* e^{2V}\phi +\tilde\phi^* e^{-2V}\tilde\phi\Big),\qquad
\end{eqnarray}

\noindent
while in the canonical normalization

\begin{eqnarray}\label{Canonical_Normalization}
&& S_{ren} =
\frac{1}{4 e^2} Z_3(\Lambda/\mu) \mbox{Re}
\int d^4x\,d^2\theta\,W_a C^{ab} W_b
+\nonumber\\
&& \qquad\qquad\qquad\qquad\qquad\qquad\quad
+ \frac{1}{4}\int d^4x\, d^4\theta\,
\Big(\phi^* e^{2V}\phi +\tilde\phi^* e^{-2V}\tilde\phi\Big).\qquad
\end{eqnarray}

\noindent
In the former case the $\beta$-function is supposed to be exhausted at
the one loop, while in the latter one it coincides with the NVSZ result.
In principle this solution is different from the one, given by Shifman
and Vainshtein. Moreover, it contradicts the results of explicit
two-loop calculations, made with DRED.

It would be natural to suppose, that in the holomorphic normalization the
$\beta$-function is exhausted at the one-loop if higher covariant
derivative regularization \cite{Slavnov,Bakeyev}, supplemented by the
Pauli-Villars, is used. This regularization is known to yield the same
result for one-loop logarithmic divergences as the dimensional
regularization (or dimensional reduction) \cite{PhysLett}. The explicit
two-loop calculations for theories, regularized by higher derivatives
(HD), were made first in \cite{hep,tmf2} for $N=1$ SUSY QED and gave a
zero two-loop contribution to the $\beta$-function defined by
\footnote{Note, that it is the $\beta$-function, that is proportional
to the trace anomaly.}

\begin{equation}\label{Beta_Definition}
\beta(\alpha) = \frac{d}{d\ln\mu}\Bigg(\frac{e^2}{4\pi}\Bigg).
\end{equation}

This result implies the absence of the anomaly puzzle in view of the
solution proposed in \cite{Arkani}. However it was not quite clear why
different regularizations give different results for the scheme independent
two-loop $\beta$-function. Actually in \cite{hep} we noted, that the using
of the HD regularization leads to a nontrivial contribution of diagrams
with insertions of one-loop counterterms, which does not exist for the
dimensional reduction. The calculations of this contribution with different
regularizations were analysed in \cite{HD_And_DRED}, where the
difference of the results for the scheme independent two-loop
$\beta$-function was attributed to the mathematical inconsistency of DRED
\cite{Siegel}, which had been pointed in \cite{Siegel2}. In particular,
the inconsistency of DRED leads to incorrect zero results for anomalies,
because DRED does not break the chiral symmetry. It is necessary to
stress an essential difference between dimensional regularization (DREG)
\cite{tHV} and DRED: DREG allows to derive the axial anomaly unambiguously
\cite{tHV}. However DREG explicitly breaks supersymmetry and is not
convenient for the calculations in supersymmetric theories. Let us note,
that anomalies can in principle be calculated with DRED. However for this
purpose it is necessary to impose mathematically inconsistent conditions
like $\mbox{tr}(AB)\ne \mbox{tr}(BA)$ \cite{Nikolai} or use some
identities between $\gamma$-matrices, which are valid only for $n>4$
\cite{Leveille}. (DRED requires that the space-time dimension $n$ should
be less than 4 \cite{Siegel}.) However such conditions can not be imposed
if the calculations are made by the supergraph technique. Hence the axial
anomaly and the Konishi anomaly, calculated with DRED, are equal to 0. As
a consequence the additional anomalous contribution, pointed in \cite{SV},
is omitted if the theory is regularized by DRED. HD regularization is
mathematically consistent and allows to calculate anomalies correctly.
In particular, the anomalous contribution to the $\beta$-function,
obtained with HD, is not equal to 0. Actually this contribution is a sum
of Feynman diagrams with insersions of counterterms on matter lines. The
sum of such diagrams is equal to 0 with DRED and agrees with the results
of \cite{SV} and \cite{Arkani} with HD regularization. After rescaling,
which converts (\ref{Holomorphic_Normalization}) into
(\ref{Canonical_Normalization}), the diagrams with insersions of
counterterms vanish, and the $\beta$-function becomes equal to the
NSVZ expression.

It is necessary to note, that although the $\beta$-function
(\ref{Beta_Definition}) is exausted by the first loop in the holomorphic
normalization, the Gell-Mann-Low function has contributions from all
orders. This contradiction is discussed in the present paper. We argue,
that if the Adler-Bardeen theorem is valid and the bare coupling constant
does not depend on $\mu$, then the generating functional depends on $\mu$
due to the rescaling anomaly and $\beta$-function (\ref{Beta_Definition})
is not related with the Gell-Mann-Low function. Therefore there is no
contradiction between the form of the Gell-Mann-Low function and the
multiplet structure of anomalies.

One more purpose of this paper is the calculation of the $\beta$-function
in the three-loop approximation. It is desirable in order to avoid some
possible errors or incorrect interpretation of the results, especially if
we take into account, that the three-loop $\beta$-function, considered as
a function of $\alpha$, is scheme-dependent. The three-loop contribution
to $\beta$-function (\ref{Beta_Definition}) is found to be 0,
and agrees with the predictions of \cite{Arkani} and \cite{HD_And_DRED}.
It is worth mentioning, that in the three-loop approximation the sum of
the diagrams without insersions of counterterms (on matter lines) for a
large number of subtraction schemes is equal to the exact $\beta$-function
(calculation with DRED gives the NSVZ $\beta$-function only after a
redefinition of the coupling constant). The sum of the diagrams with
insersions of counterterms in two- and three-loop approximations agrees
with the exact expression found in \cite{HD_And_DRED}, and cancels the
other two- and three-loop contributions.

The paper is organized as follows:

In section \ref{Section_SUSY_QED} we consider $N=1$ SUSY QED and
regularize it by higher derivatives. The three-loop $\beta$-function and
its relation with two-loop anomalous dimension are analysed in section
\ref{Section_Three_Loop}. In particular, the three-loop contribution to
the $\beta$-function is found to be 0. In section \ref{Section_HD_And_DRED}
we explain why the results are different from those obtained with DRED.
The anomaly puzzle is considered in section \ref{Section_Anomaly_Puzzle}.
Section \ref{Section_Conclusion} contains some concluding remarks.
The details of the calculations are presented in appendixes. Appendix
\ref{Appendix_Diagrams} contains expressions for various groups of
Feynman diagrams. The calculations of the corresponding contributions are
made in appendix \ref{Appendix_Relation} and the most useful three-loop
integrals are analysed in appendix \ref{Appendix_Integrals}.


\section{$N=1$ supersymmetric electrodynamics and higher derivative
regularization.}
\label{Section_SUSY_QED}
\hspace{\parindent}

$N=1$ supersymmetric electrodynamics is described by the following action:

\begin{equation}\label{SQED_Action}
S_0 = \frac{1}{4 e^2} \mbox{Re}\int d^4x\,d^2\theta\,W_a C^{ab} W_b
+ \frac{1}{4}\int d^4x\, d^4\theta\,
\Big(\phi^* e^{2V}\phi +\tilde\phi^*
e^{-2V}\tilde\phi\Big).
\end{equation}

\noindent
Here $\phi$ and $\tilde\phi$ are chiral superfields

\begin{eqnarray}\label{Phi_Superfield}
&& \phi(y,\theta) = \varphi(y) + \bar\theta (1+\gamma_5) \psi(y)
+ \frac{1}{2}\bar\theta (1+\gamma_5)\theta f(y);\nonumber\\
&&\tilde \phi(y,\theta) = \tilde \varphi(y)
+ \bar\theta (1+\gamma_5) \tilde \psi(y)
+ \frac{1}{2}\bar\theta (1+\gamma_5)\theta \tilde f(y),
\end{eqnarray}

\noindent
where $y^\mu = x^\mu + i\bar\theta\gamma^\mu\gamma_5\theta/2$.
Two Majorana spinors $\psi$ and $\tilde\psi$ form one Dirac spinor

\begin{equation}\label{Psi_Definition}
\Psi = \frac{1}{\sqrt{2}}\Big((1+\gamma_5)\psi+(1-\gamma_5)\tilde\psi\Big).
\end{equation}

\noindent
$V$ in (\ref{SQED_Action}) is a real superfield

\begin{eqnarray}\label{V_Superfield}
&& V(x,\theta) = C(x)+i\sqrt{2}\bar\theta\gamma_5\xi(x)
+\frac{1}{2}(\bar\theta\theta)K(x)
+\frac{i}{2}(\bar\theta\gamma_5\theta)H(x)
+\frac{1}{2}(\bar\theta \gamma^\mu \gamma_5\theta) A_\mu(x)
+\nonumber\\
&& + \sqrt{2} (\bar\theta\theta) \bar\theta
\Big(i\gamma_5\chi(x)
+\frac{1}{2}\gamma^\mu\gamma_5\partial_\mu\xi(x)\Big)
+ \frac{1}{4} (\bar\theta\theta)^2 \Big(D(x)
-\frac{1}{2}\partial^2 C(x)\Big),
\end{eqnarray}

\noindent
where, in particular, $A_\mu$ is an Abelian gauge field. The superfield
$W_a$ in the Abelian case is defined by

\begin{equation}
W_a = \frac{1}{16} \bar D (1-\gamma_5) D\Big[(1+\gamma_5)D_a V\Big],
\end{equation}

\noindent
where

\begin{equation}
D = \frac{\partial}{\partial\bar\theta} - i\gamma^\mu\theta\,\partial_\mu
\end{equation}

\noindent
is a supersymmetric covariant derivative.

In order to regularize model (\ref{SQED_Action}) by HD its action should
be modified as follows:

\begin{eqnarray}\label{Regularized_SQED_Action}
&& S_0 \to S = S_0 + S_{\Lambda}
=\vphantom{\frac{1}{2}}\nonumber\\
&&\qquad
= \frac{1}{4 e^2} \mbox{Re}\int d^4x\,d^2\theta\,W_a C^{ab}
\Big(1+ \frac{\partial^{2n}}{\Lambda^{2n}}\Big) W_b
+\nonumber\\
&&\qquad\qquad\qquad\qquad\qquad\qquad
+ \frac{1}{4}\int d^4x\, d^4\theta\,
\Big(\phi^* e^{2V}\phi +\tilde\phi^* e^{-2V}\tilde\phi\Big).\qquad
\end{eqnarray}

\noindent
Note, that in the Abelian case the superfield $W^a$ is gauge invariant,
so the higher derivative term contains usual derivatives.

The quantization of (\ref{Regularized_SQED_Action}) can be made by using
standard technique described in \cite{West} and is not considered here.
It only needs mentioning that the gauge invariance was fixed by adding

\begin{equation}
S_{gf} = - \frac{1}{64 e^2}\int d^4x\,d^4\theta\,
\Bigg(V D^2 \bar D^2
\Big(1 + \frac{\partial^{2n}}{\Lambda^{2n}}\Big) V
+ V \bar D^2 D^2
\Big(1+ \frac{\partial^{2n}}{\Lambda^{2n}}\Big) V\Bigg),
\end{equation}

\noindent
where

\begin{equation}
D^2 \equiv \frac{1}{2} \bar D (1+\gamma_5)D;\qquad
\bar D^2 \equiv \frac{1}{2}\bar D (1-\gamma_5) D.
\end{equation}

\noindent
After adding such terms the free part of the action for the
superfield $V$ is written in the simplest form

\begin{equation}
S_{gauge} + S_{gf} = \frac{1}{4 e^2}\int d^4x\,d^4\theta\,
V\partial^2 \Big(1+ \frac{\partial^{2n}}{\Lambda^{2n}}\Big) V.
\end{equation}

\noindent
In the Abelian case diagrams containing ghost loops are missing.

The superficial degree of divergence for the model
(\ref{Regularized_SQED_Action}) is (see e.f. \cite{hep})

\begin{equation}\label{Degree_Of_Divergence}
\omega_\Lambda = 2 - 2n (L-1) - E_\phi (n+1),
\end{equation}

\noindent
where $L$ is a number of loops and $E_\phi$ is a number of external
$\phi$-lines. According to (\ref{Degree_Of_Divergence}) divergences
remain in one-loop diagrams even for $n\ge 2$. In order to regularize
these divergences it is necessary to insert Pauli-Villars determinants
\cite{Slavnov_Book} into the generating functional. Due to the
supersymmetric gauge invariance

\begin{equation}
V \to V - \frac{1}{2}(A+A^+);
\qquad \phi\to e^{A}\phi;\qquad \tilde\phi\to e^{-A} \tilde\phi,
\end{equation}

\noindent
where $A$ is an arbitrary chiral scalar superfield, the renormalized
action can be written as

\begin{eqnarray}\label{Renormalized_Action}
&& S_{ren} =
\frac{1}{4 e^2} Z_3(\Lambda/\mu) \mbox{Re}\int d^4x\,d^2\theta\,W_a C^{ab}
\Big(1+ \frac{\partial^{2n}}{\Lambda^{2n}}\Big) W_b
+\nonumber\\
&& \qquad\qquad\qquad\qquad\qquad
+ Z(\Lambda/\mu)\frac{1}{4}\int d^4x\, d^4\theta\,
\Big(\phi^* e^{2V}\phi +\tilde\phi^* e^{-2V}\tilde\phi\Big).\qquad
\end{eqnarray}

\noindent
Hence the generating functional is

\begin{eqnarray}\label{Modified_Z}
&& Z = \int DV\,D\phi\,D\tilde \phi\,
\prod\limits_i \Big(\det PV(V,M_i)\Big)^{c_i}
\exp\Bigg\{i\Bigg[\frac{1}{4 e^2} \int d^4x\,d^4\theta\, V\partial^2
\Big(1+ \frac{\partial^{2n}}{\Lambda^{2n}}\Big) V
-\nonumber\\
&& - \frac{1}{4 e^2} \Big(Z_3(\Lambda/\mu)-1\Big) \int d^4x\,d^4\theta\,
V \Pi_{1/2}\partial^2
\Big(1+ \frac{\partial^{2n}}{\Lambda^{2n}}\Big) V
+\nonumber\\
&& \qquad\qquad\qquad\qquad\qquad\qquad\qquad\qquad
+ \frac{1}{4} Z(\Lambda/\mu) \int d^4x\,d^4\theta\,
\Big(\phi^* e^{2V}\phi
+ \tilde\phi^* e^{-2V}\tilde\phi \Big)
+\nonumber\\
&&
+ \int d^4x\,d^4\theta\,J V
+ \int d^4x\,d^2\theta\, \Big(j\,\phi + \tilde j\,\tilde\phi \Big)
+ \int d^4x\,d^2\bar\theta\,
\Big(j^*\phi^* + \tilde j^* \tilde\phi^* \Big)\Bigg]\Bigg\},
\end{eqnarray}

\noindent
where

\begin{eqnarray}\label{PV_Determinants}
&& \Big(\det PV(V,M)\Big)^{-1} = \int D\Phi\,D\tilde \Phi\,
\exp\Bigg\{i\Bigg[ Z(\Lambda/\mu) \frac{1}{4} \int d^4x\,d^4\theta\,
\Big(\Phi^* e^{2V}\Phi
+\qquad\nonumber\\
&& + \tilde\Phi^* e^{-2V}\tilde\Phi \Big)
+ \frac{1}{2}\int d^4x\,d^2\theta\, M \tilde\Phi \Phi
+ \frac{1}{2}\int d^4x\,d^2\bar\theta\, M \tilde\Phi^* \Phi^*
\Bigg]\Bigg\},
\end{eqnarray}

\noindent
and the coefficients $c_i$ satisfy equations

\begin{equation}
\sum\limits_i c_i = 1;\qquad \sum\limits_i c_i M_i^2 = 0.
\end{equation}

\noindent
Below we assume, that $M_i = a_i\Lambda$, where $a_i$ are some
constants. The insertion of Pauli-Villars determinants allows us to
cancel remaining divergences in all one-loop diagrams, including diagrams
with insertions of counterterms. Later we will show, that the divergencies
in the sum of two- and three-loop diagrams with Pauli-Villars loops
cancel each other. Therefore, for diagrams with loops of Pauli-Villars
fields it is unnecessary to introduce any other regularization.

In our notations the generating functional for connected Green functions
is defined by

\begin{equation}\label{W}
W = - i\ln Z,
\end{equation}

\noindent
and an effective action is obtained by making a Legendre transformation:

\begin{equation}\label{Gamma}
\Gamma = W - \int d^4x\,d^4\theta\,J V
- \int d^4x\,d^2\theta\, \Big(j\,\phi + \tilde j\,\tilde\phi \Big)
- \int d^4x\,d^2\bar\theta\,
\Big(j^*\phi^* + \tilde j^* \tilde\phi^* \Big),
\end{equation}

\noindent
where $J$, $j$ and $\tilde j$ is to be eliminated in terms of
$V$, $\phi$ and $\tilde\phi$, through solving equations

\begin{equation}
V = \frac{\delta W}{\delta J};\qquad
\phi = \frac{\delta W}{\delta j};\qquad
\tilde\phi = \frac{\delta W}{\delta\tilde j}.
\end{equation}

After obtaining $S_{ren}$, it is possible to find the $\beta$-function
and the anomalous dimension, which in our notations are defined by

\begin{equation}\label{Beta_Gamma_Definition}
\beta = \frac{d}{d\ln\mu}\Bigg(\frac{e^2}{4\pi}\Bigg);
\qquad\quad
\gamma = \frac{d\ln Z}{d\ln\mu}.
\end{equation}

\noindent
(We assume, that the bare coupling constant $e_0$, defined by

\begin{equation}\label{Bare_Coupling_Constant}
\frac{1}{e_0^2} = \frac{1}{e^2} Z_3(\Lambda/\mu),
\end{equation}

\noindent
does not depend on $\mu$. Hence the renormalized coupling constant $e$
depends on $\mu$.) It is easy to see \cite{Adler_Collins}, that the trace
anomaly is proportional to $\beta$-function (\ref{Beta_Gamma_Definition}).

The $\beta$-function and anomalous dimension, given by
(\ref{Beta_Gamma_Definition}), which are considered as functions of
$e$, are changed at the simultanious redefinition of the renormalized
coupling $e$ and the renormalization constant $Z_3$, provided
$e_0=\mbox{const}$. In other words they depend on the renormalization
scheme. If $\beta$ and $\gamma$ are expanded in powers of $e^2$, then
the coefficients of the $\beta$-function and anomalous dimension become
scheme-dependent starting from the three- and two-loop approximation
respectively.

Note, that it is possible to use another definition of the
$\beta$-function. Let us consider transversal part of the two-point Green
function for the gauge field:

\begin{eqnarray}\label{Two_Point_Function}
&& \Pi_{1/2}\,
\int d^4x\,d^4y\,\frac{\delta^2 \Gamma}{\delta V_x\,\delta V_y}
\Bigg|_{J=0} \exp\Big(i p_\mu x^\mu + i q_\mu y^\mu \Big)
=\nonumber\\
&&\qquad\qquad\qquad\qquad
= \frac{1}{8\pi} (2\pi)^4 \delta^4\Big(p + q\Big)\,
p^2 \Pi_{1/2} \delta^4(\theta_x-\theta_y)\,d^{-1}(\alpha,\mu/p),\qquad
\end{eqnarray}

\noindent
where

\begin{equation}
\Pi_{1/2} \equiv - \frac{1}{16 \partial^2} D^a \bar D^2 C_{ab} D^b.
\end{equation}

\noindent
Then it is possible to define Gell-Mann-Low function

\begin{equation}\label{GL_Function}
\tilde\beta\Big(d(\alpha,x)\Big) \equiv
- x\frac{\partial}{\partial x} d(\alpha,x).
\end{equation}

\noindent
Taking into account, that the effective action should not depend
on the normalization point $\mu$ and differentiating equation
(\ref{Two_Point_Function}) over $\ln\mu$ we obtain

\begin{equation}
0 = \tilde\beta\Big(d(\alpha,x)\Big)
- \beta(\alpha) \frac{\partial}{\partial\alpha} d(\alpha,x).
\end{equation}

\noindent
In particular at $x=1$ we have

\begin{equation}\label{GL_And_Beta}
\tilde\beta(\tilde\alpha) = \beta(\alpha)\frac{d\tilde\alpha}{d\alpha},
\end{equation}

\noindent
where $\tilde\alpha\equiv d(\alpha,1)$. Therefore, if the generating
functional does not depend on $\mu$, then both definitions of the
$\beta$-function are equivalent.

In order to find the $\beta$-function and the anomalous dimension it is
necessary to calculate all 1PI graphs in the considered approximation.
The expressions for them are constructed in accordance with Feynman rules,
which can be formulated as follows:

1. External lines give the integration

\begin{equation}
\prod\limits_{E}
\int \frac{d^4p_{{}_{E_V}}}{(2\pi)^4} V(p_{{}_{E_V}})
\int \frac{d^4p_{{}_{E_\phi}}}{(2\pi)^4} \phi(p_{{}_{E_\phi}})
\cdot \ldots\cdot
(2\pi)^4 \delta\Big(\sum\limits_{E} p_{{}_E}\Big),
\end{equation}

\noindent
where $E$ runs over external momenta.

2. Propagator of the superfield $V$ is

\begin{equation}
\frac{8e^2}{(k^2+i0) \Big(1+(-1)^n k^{2n}/\Lambda^{2n}\Big)} \,
\delta^4(\theta_1-\theta_2).
\end{equation}

3. Massless $\phi-\phi^*$ and $\tilde\phi-\tilde\phi^*$ propagators
are

\begin{equation}
-\frac{1}{16 (k^2+i0)}\bar D^2 D^2 \delta^4(\theta_1-\theta_2).
\end{equation}

\noindent
(Note that the considered action is quadratic in matter superfields
and Feynman rules can be simplified in comparison with, say,
Wess-Zumino model.)

4. Pauli-Villars fields are present only in the closed loops. Each
internal line $\Phi-\Phi^*$ or $\tilde\Phi-\tilde\Phi^*$ corresponds to

\begin{equation}
- \frac{1}{16(k^2-M_i^2+i0)}\, \bar D^2 D^2 \delta^4(\theta_1-\theta_2),
\end{equation}

\noindent
and each internal line $\Phi-\tilde\Phi$ or $\Phi^*-\tilde\Phi^*$ -- to

\begin{equation}
\frac{M_i}{4(k^2-M_i^2+i0)}\,\bar D^2 \delta^4(\theta_1-\theta_2)
\quad\mbox{and}\quad
\frac{M_i}{4(k^2-M_i^2+i0)}\, D^2 \delta^4(\theta_1-\theta_2)
\end{equation}

\noindent
respectively. For each loop of Pauli-Villars fields it is necessary to
introduce ${\displaystyle - \sum\limits_i c_i}$.

5. Each loop yields an integration over a loop momentum
${\displaystyle \int \frac{d^4k}{(2\pi)^4}}$.

6. Each vertex gives ${\displaystyle \int d^4\theta}$.

7. There are usual combinatoric factors, which can be found from
the generating functional (\ref{Modified_Z}).


\section{Three-loop $\beta$-function.}
\hspace{\parindent}
\label{Section_Three_Loop}

\begin{figure}[h]
\hspace*{3cm}
\epsfxsize7.5truecm\epsfbox{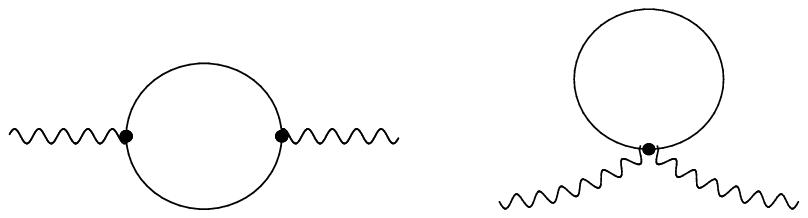}
\caption{One-loop graphs for calculating $\Delta\Gamma_V$.}
\label{Figure_Beta_Diagrams_1Loop}
\end{figure}

\begin{figure}[h]
\hspace*{1.9cm}
\epsfxsize11.0truecm\epsfbox{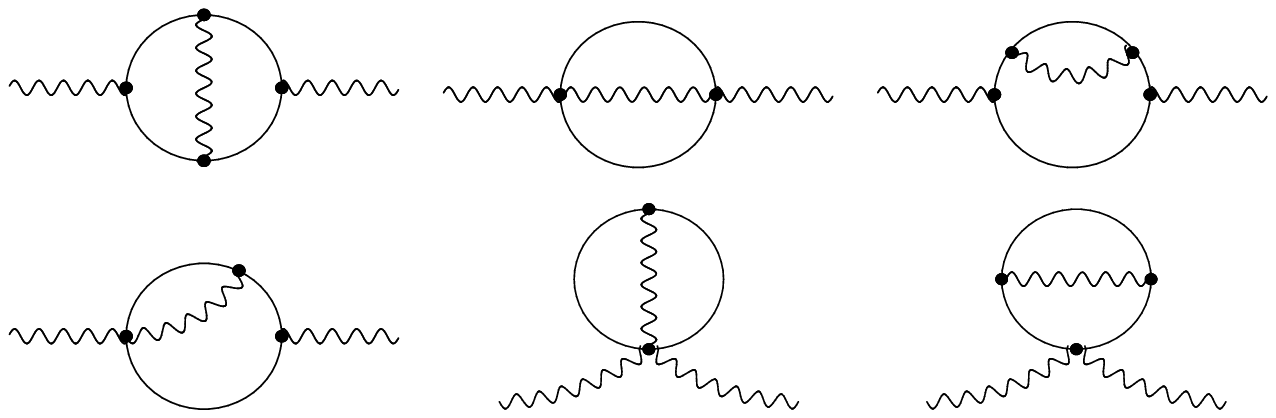}
\caption{Two-loop graphs for calculating $\Delta\Gamma_V$.}
\label{Figure_Beta_Diagrams_2Loop}
\end{figure}

\begin{figure}[h]
\hspace*{-7mm}
\epsfxsize16.5truecm\epsfbox{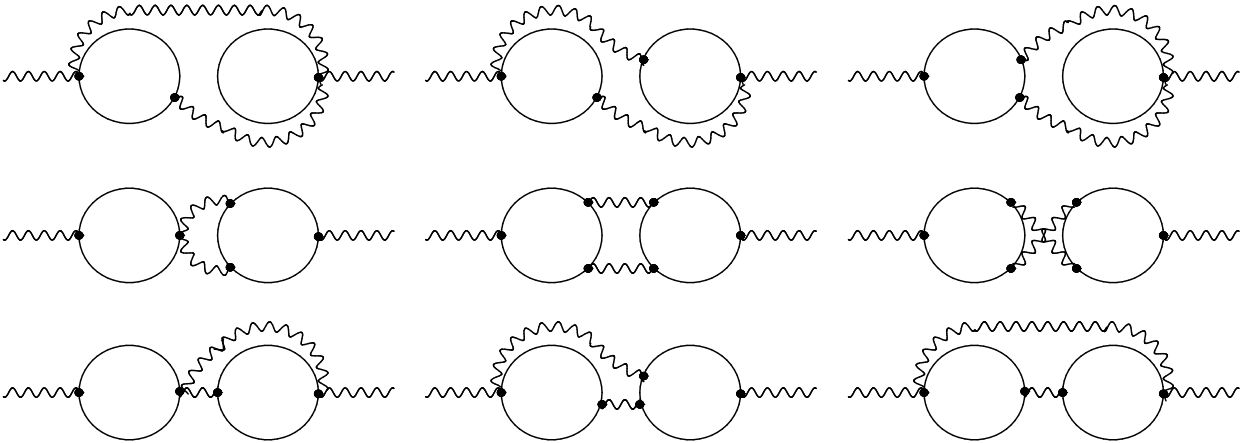}
\caption{Three-loop graphs with two loops of the matter fields.
In these diagrams external lines are attached to both loops.}
\label{Figure_Beta_Diagrams_OO}
\end{figure}

\begin{figure}[h]
\epsfxsize15.0truecm\epsfbox{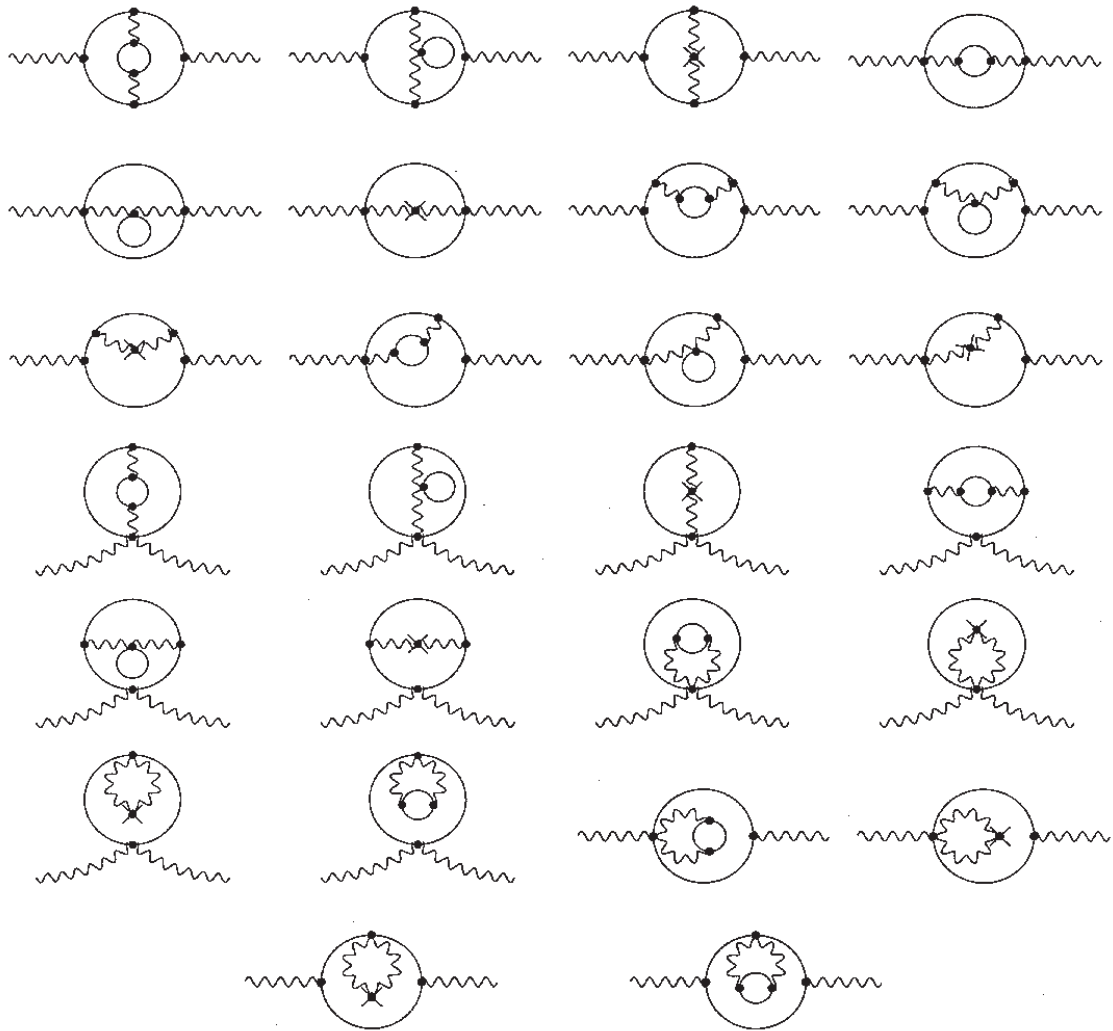}
\caption{Contribution to $\Delta\Gamma_V$, corresponding to corrections
to the photon propagator.}
\label{Figure_Beta_Diagrams_Oo}
\end{figure}

\begin{figure}[p]
\epsfxsize15.0truecm\epsfbox{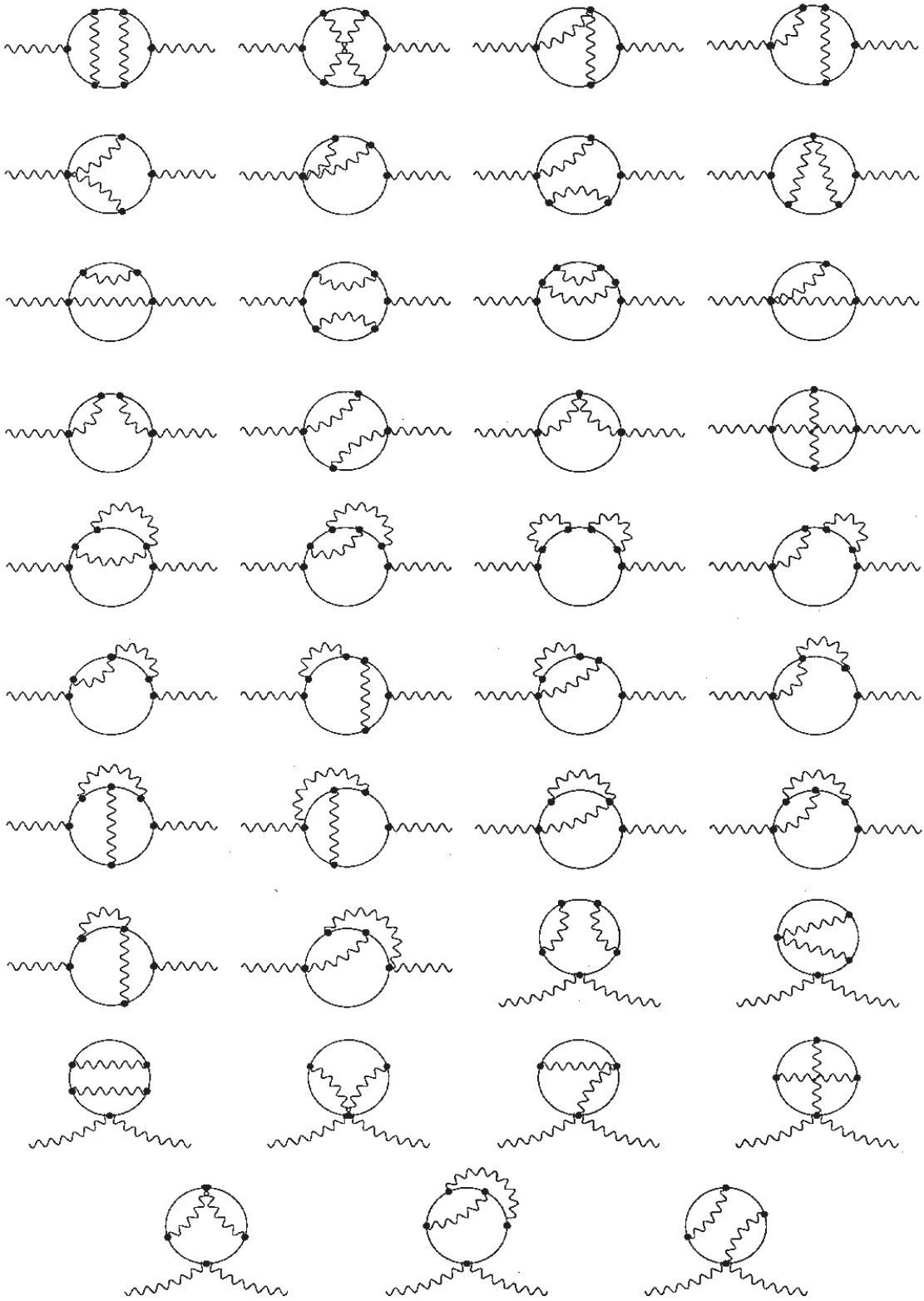}
\caption{Three-loop graphs with a single loop of matter superfield.}
\label{Figure_Beta_Diagrams_O}
\end{figure}

\begin{figure}[h]
\epsfxsize15.0truecm\epsfbox{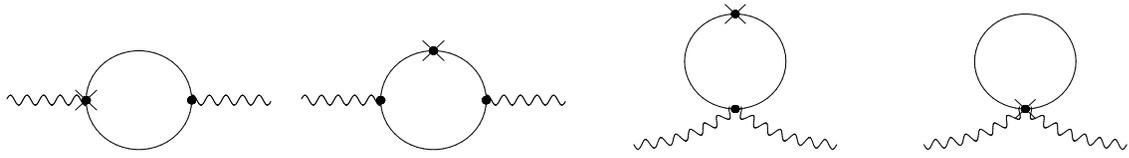}
\caption{Diagrams, containing an insersion of two-loop counterterms
on the matter lines.}
\label{Figure_Beta_Diagrams_X}
\end{figure}

\begin{figure}[h]
\epsfxsize15.0truecm\epsfbox{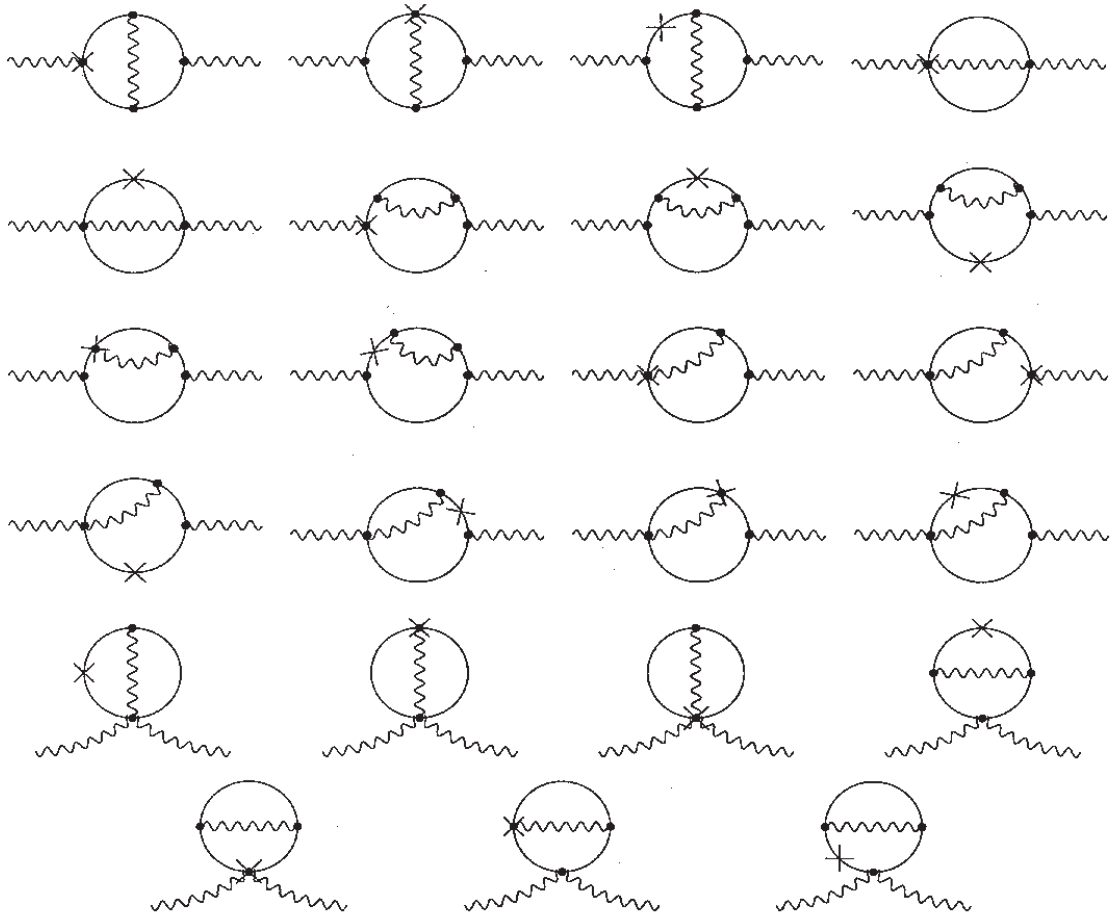}
\caption{Diagrams containing an insersion of one-loop counterterms.}
\label{Figure_Beta_Diagrams_X2}
\end{figure}

\begin{figure}[h]
\hspace*{1.9cm}
\epsfxsize11.0truecm\epsfbox{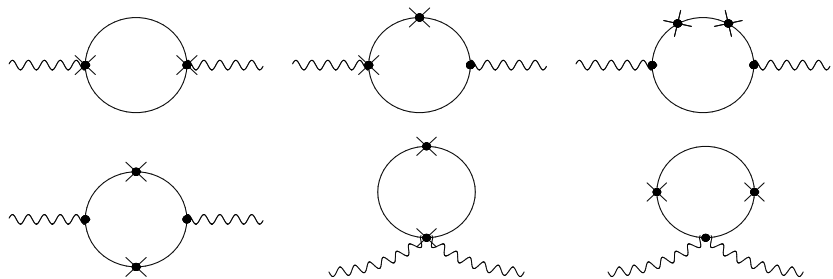}
\caption{Feynman diagrams with two insersions of counterterms.}
\label{Figure_Beta_Diagrams_XX}
\end{figure}

The diagrams contributing to the three-loop $\beta$-function are presented
in Fig. \ref{Figure_Beta_Diagrams_1Loop}~--~\ref{Figure_Beta_Diagrams_XX}.
Note, that each internal matter loop in these diagrams can correspond to
$\phi$ and $\tilde\phi$ fields or to Pauli-Villars fields. We devided
Feynman diagrams into some groups and presented the expressions for all
these groups in appendix \ref{Appendix_Diagrams}. Then the three-loop
correction to the effective action can be written as

\begin{eqnarray}\label{Three_Loop_Effective_Action_V}
&& \Delta\Gamma^{(3)}_{V} = \mbox{Re} \int d^2\theta\,
\frac{d^4p}{(2\pi)^4} W_a(p) C^{ab} W_b(-p)
\times\nonumber\\
&&\qquad\qquad
\times
\Big(f_{1-loop} + f_{O} + f_{O_{PV}} + f_{Oo} + f_{Oo_{PV}} + f_{OO}
+ f_{XX} + f_{X} + f_{X2} \Big),\qquad
\end{eqnarray}

\noindent
where

$f_{1-loop}$ is a contribution of one-loop diagrams, presented in
Fig. \ref{Figure_Beta_Diagrams_1Loop};

$f_{O}$ is a contribution of diagrams, presented in Fig.
\ref{Figure_Beta_Diagrams_O}, containing a single loop of the
superfields $\phi$ and $\tilde\phi$;

$f_{O_{PV}}$ is a contribution of the same diagrams, having a loop of
Pauli-Villars fields;

$f_{Oo}$ is a contribution of the two-loop diagrams, presented in Fig.
\ref{Figure_Beta_Diagrams_2Loop} (without Pauli-Villars fields) and
the three-loop diagrams, presented in Fig. \ref{Figure_Beta_Diagrams_Oo},
which contain an internal loop of matter superfields or an insersion
of one-loop counterterms on the photon line and external loop of
$\phi$ and $\tilde\phi$ superfields;

$f_{Oo_{PV}}$ is a contribution of the same diagrams with the external
loop of Pauli-Villars fields.

$f_{OO}$ is a contribution of the other diagrams with two loops of
the matter superfields, presented in Fig. \ref{Figure_Beta_Diagrams_OO};

$f_{X}$ is a contribution of diagrams with an insersion of two-loop
counterterms, presented in Fig. \ref{Figure_Beta_Diagrams_X};

$f_{XX}$ is a contribution of diagrams with two insersions of one-loop
counterterms, presented in Fig. \ref{Figure_Beta_Diagrams_XX};

$f_{X2}$ is a contribution of diagrams with an insersion of one-loop
counterterms, presented in Fig. \ref{Figure_Beta_Diagrams_X2}.

The explicit expressions for all these contributions were found
by means of calculating of the corresponding Feynman diagrams. To check
correctness of these calculations we verified the cancellation of
noninvariant terms, proportional to

\begin{equation}
\int \frac{d^4p}{(2\pi)^4}\,d^4\theta\,V(p,\theta)\,V(-p,\theta).
\end{equation}

\noindent
The results, presented in appendix \ref{Appendix_Diagrams}, are analysed
in appendix \ref{Appendix_Relation}. Let us briefly discuss them:

1. $f_{OO} = 0$, because the substitution $\phi\leftrightarrow\tilde\phi$
in a loop changes the sign of a diagram. Indeed, in this case the diagrams,
having the same superfields in both loops ($\phi$ and $\phi$ or
$\tilde\phi$ and $\tilde\phi$), are cancelled by diagrams with loops of
different superfields ($\phi$ and $\tilde\phi$). For the diagrams with
Pauli-Villars fields the result is the same, but its derivation is more
complicated.

2. The sum of $f_{XX}$, $f_X$ and $f_{X2}$ agrees with the exact
expression for the sum of diagrams with insersions of counterterms

\begin{equation}\label{Delta_GammaV}
- \ln Z\,\frac{1}{16\pi^2}\,
\mbox{Re}\int d^4x\,d^2\theta\,W_a C^{ab} W_b +\mbox{finite terms},
\end{equation}

\noindent
found in \cite{HD_And_DRED}. The corresponding contribution to
$\beta$-function in the considered approximation is

\begin{equation}\label{Delta_Beta}
\Delta\beta = \frac{\alpha^2}{\pi}\gamma(\alpha).
\end{equation}

According to the results of the one-loop calculations and the predictions
of the renormgroup (see e.f. \cite{hep}) the constant $Z$ is given by

\begin{eqnarray}\label{Constant_Z}
&& Z(\Lambda/\mu) = 1 + \frac{\alpha}{\pi}\,
\Big(\ln \frac{\Lambda}{\mu}+g_1\Big)
+\nonumber\\
&& \qquad\qquad
+ \frac{\alpha^2}{\pi^2}\,\Big(\ln^2 \frac{\Lambda}{\mu}
+ g_1 \ln\frac{\Lambda}{\mu}\Big)
- \gamma_2\, \alpha^2\, \Big(\ln \frac{\Lambda}{\mu}
+ g_2\Big)+ O(\alpha^3).\qquad
\end{eqnarray}

\noindent
Here $\gamma_2\alpha^2$ is a two-loop contribution to the anomalous
dimension and we assume, that at the one-loop the counterterms are

\begin{eqnarray}
&& \Delta S = - \frac{1}{16\pi^2} \Big(\ln\frac{\Lambda}{\mu}+b_1\Big)
\mbox{Re}\int d^4x\,d^2\theta\,W_a C^{ab}
\Big(1+ \frac{\partial^{2n}}{\Lambda^{2n}}\Big) W_b
+\nonumber\\
&&\qquad\qquad\qquad\qquad\qquad
+ \frac{e^2}{16\pi^2}\Big(\ln\frac{\Lambda}{\mu}+g_1\Big)
\int d^4x\,d^4\theta\,
\Big(\phi^* e^{2V}\phi +\tilde\phi^* e^{-2V}\tilde\phi\Big),\qquad
\end{eqnarray}

\noindent
where $b_1$, $g_1$ and $g_2$ are arbitrary finite constants,
which define a subtraction scheme. \footnote{It is convenient to include
the higher derivative term in counterterms, because its presence simplifies
an expression for the two-loop anomalous dimension \cite{tmf2}.
In principle, this term is not essential and can be omitted.}

3. $f_{Oo_{PV}}$ and $f_{O_{PV}}$ are finite and do not contribute to
the divergent part of the effective action. This means, that
the sum of all diagrams with Pauli-Villars loops is finite, although
there are divergences in some of such graphs. However, Pauli-Villars
regularization always assumes the existance of divergent diagrams and
the cancellation of the divergences between different graphs. Therefore,
in the considered case it is not necessary to introduce any more
regularization.

4. The analysis of $f_{Oo}$ and $f_{O}$ is rather involved, because the
corresponding integrals are very complicated. Each of these integrals
depends on $\Lambda/p$ and is the sum of a third degree polynomial in
$\ln\Lambda/p$ and a function, finite at $\Lambda\to\infty$ (or
equivalently at $p\to 0$)

\begin{equation}\label{f_Expansion}
f(\Lambda/p) = f_3 \ln^3 \frac{\Lambda}{p} + f_2 \ln^2\frac{\Lambda}{p}
+ f_1 \ln\frac{\Lambda}{p} + f_0 + o(1).
\end{equation}

\noindent
Let us assume, that the limit

\begin{equation}\label{Limit}
\Lambda \frac{d}{d\Lambda} f(\Lambda/p)\Bigg|_{p=0}
\end{equation}

\noindent
exists. Then $f_3=0$ and $f_2=0$, while the considered limit is $f_1$.

In appendix \ref{Appendix_Relation} we prove, that for $f_{Oo}$ and $f_O$
the limit (\ref{Limit}) exists and

\begin{eqnarray}\label{Remaining_Integral}
\Lambda \frac{d}{d\Lambda} \Big(f_{Oo}+f_O\Big)\Bigg|_{p=0}
= \frac{1}{16\pi^2} \Lambda \frac{d}{d\Lambda}
\Big(z_1+z_2+\frac{1}{2}z_1^2\Big)\Bigg|_{p=0},
\end{eqnarray}

\noindent
where

\begin{eqnarray}\label{Z1_Definition}
&& z_1 \equiv \int \frac{d^4k}{(2\pi)^4}\,
\frac{2e^2}{k^2 (k+p)^2 \Big(1+k^{2n}/\Lambda^{2n}\Big)};\\
\label{Z2_Definition}
&& z_2 \equiv
- \int \frac{d^4k}{(2\pi)^4}\,\frac{d^4l}{(2\pi)^4}\,
\frac{4 e^4}{k^2 l^2 (k+p)^2 (l+p)^2
\Big(1+ k^{2n}/\Lambda^{2n}\Big)
\Big(1+ l^{2n}/\Lambda^{2n}\Big)}
-\nonumber\\
&& - \int \frac{d^4k}{(2\pi)^4}\,\frac{d^4l}{(2\pi)^4}\,
\frac{4 e^4}{\displaystyle k^2 l^2 (l+p)^2 (k+l+p)^2
\Big(1+ k^{2n}/\Lambda^{2n}\Big)
\Big(1+ l^{2n}/\Lambda^{2n}\Big)}
+\nonumber\\
&& + \int \frac{d^4k}{(2\pi)^4}\,\frac{d^4l}{(2\pi)^4}\,
\frac{4 e^4 (k+l+2p)^2}{k^2 (k+p)^2 l^2 (l+p)^2 (k+l+p)^2
\Big(1+ k^{2n}/\Lambda^{2n}\Big)
\Big(1+ l^{2n}/\Lambda^{2n}\Big)}
-\quad\nonumber\\
&& - \int \frac{d^4k}{(2\pi)^4}
\,\frac{4 e^4}{k^2 (k+p)^2 \Big(1+k^{2n}/\Lambda^{2n}\Big)^2}
\Bigg(\int \frac{d^4l}{(2\pi)^4}\, \frac{1}{l^2 (k+l)^2}
-\\
&& - \sum\limits_i c_i
\int \frac{d^4l}{(2\pi)^4}\,\frac{1}{(l^2+M_i^2)\Big((k+l)^2+M_i^2\Big)}
- \frac{1}{8\pi^2}\Big(\ln\frac{\Lambda}{\mu}+b_1\Big)
\Big(1+k^{2n}/\Lambda^{2n}\Big)
\Bigg).\qquad\nonumber
\end{eqnarray}

\noindent
From (\ref{Remaining_Integral}), (\ref{Z1_Definition}) and
(\ref{Z2_Definition}) we see, that the integral over three loop momenta
is reduced to the integral over two loop momenta. It is very nontrivial,
that can be seen from the calculations, done in appendixes
\ref{Appendix_Relation} and \ref{Appendix_Integrals}. In our opinion
these facts confirm the correctness of the obtained results.

Note, that $z_1$ and $z_2$ are present in the two-loop two-point
Green function for the matter superfield \cite{tmf2}:

\begin{eqnarray}\label{Renormalized_Gamma_2}
&& \Gamma^{(2)}_\phi =
\frac{1}{4}\int d^4\theta\,\frac{d^4p}{(2\pi)^4}\,
\Big(\phi^*(p,\theta)\,\phi(-p,\theta)
+ \tilde\phi^*(p,\theta)\,\tilde\phi(-p,\theta)\Big)
\Bigg\{ 1-z_1-z_2
-\qquad\nonumber\\
&& - \frac{e^2}{4\pi^2} z_1 \Big(\ln\frac{\Lambda}{\mu}+g_1\Big)
+ \frac{e^2}{4\pi^2} \Big(\ln\frac{\Lambda}{\mu}+g_1\Big)
+\mbox{two-loop counterterms}
\Bigg\}.
\end{eqnarray}

\noindent
This expression can be formally written as

\begin{eqnarray}\label{Abstract_Gamma}
&& \Gamma^{(2)}_\phi =
\frac{1}{4}\int d^4\theta\,\frac{d^4p}{(2\pi)^4}\,
\Big(\phi^*(p,\theta)\,\phi(-p,\theta)
+ \tilde\phi^*(p,\theta)\,\tilde\phi(-p,\theta)\Big)
\times\nonumber\\
&& \qquad\qquad\qquad\qquad\quad
\times
\Bigg\{ 1-z_1-z_2 - \hat\Delta z_1\,z_1
+ \hat\Delta z_1 + \hat\Delta z_2
+ \hat\Delta \Big(\hat\Delta z_1\,z_1\Big)
\Bigg\},\qquad
\end{eqnarray}

\noindent
where the operator $\hat\Delta$ is constructed as follows:
If $f$ is a function of $\Lambda/p$ and $\Lambda/\mu$, then
by definition $\hat \Delta f$ is a counterterm, which cancels a
divergence of the function $f$. For example,

\begin{equation}
\hat\Delta z_1 = \frac{\alpha}{\pi}\Big(\ln\frac{\Lambda}{\mu} + g_1\Big).
\end{equation}

\noindent
Below we will assume, that the operator $\hat\Delta$ is linear.
(In a general case this operator can be nonlinear).

From (\ref{Abstract_Gamma}) we see, that the two-loop renormalization
constant for the matter superfield is given by

\begin{equation}
Z = 1+\hat\Delta z_1+\hat\Delta z_2+\hat\Delta(\hat\Delta z_1\, z_1)
+O(\alpha^3),
\end{equation}

\noindent
so that

\begin{eqnarray}
&& \ln Z = \hat\Delta z_1 + \hat\Delta z_2 + \hat\Delta(\hat\Delta z_1\,z_1)
- \frac{1}{2}\Big(\hat\Delta z_1\Big)^2 + O(\alpha^3)
= \hat\Delta z_1 + \hat\Delta z_2
-\nonumber\\
&& - \frac{1}{2}\hat\Delta\Big(z_1 - \hat\Delta z_1\Big)^2
+ \frac{1}{2}\hat\Delta\Big(z_1^2\Big)
+ \frac{1}{2}\hat\Delta\Big((\hat\Delta z_1)^2\Big)
- \frac{1}{2}(\hat\Delta z_1)^2 + O(\alpha^3).\qquad
\end{eqnarray}

\noindent
Taking into account, that due to the definition of $\hat\Delta$
the expressions $(z_1 - \hat\Delta z_1)^2$ and
$\hat\Delta\Big((\hat\Delta z_1)^2\Big) - (\hat\Delta z_1)^2$
are finite, $\ln Z$ can be presented as

\begin{equation}\label{Ln_Z_Final}
\ln Z = \hat\Delta \Big(z_1 + z_2 + \frac{1}{2} z_1^2\Big)
+\mbox{finite terms}.
\end{equation}

\noindent
Then the sum of diagrams, defining two- and three-loop contributions
to the $\beta$-function for $N=1$ SUSY QED, for subtraction schemes,
corresponding to any linear operator $\hat\Delta$ can be written in the
following form:

\begin{eqnarray}
&& \Delta\Gamma^{(3)}_{V} = \frac{1}{16\pi^2}\mbox{Re} \int d^2\theta\,
\frac{d^4p}{(2\pi)^4} W_a(p) C^{ab} W_b(-p)
\times\nonumber\\
&&\qquad\qquad\qquad
\times
\Bigg(
z_1 + z_2 + \frac{1}{2} z_1^2
- \hat\Delta\Big(z_1+z_2+\frac{1}{2}z_1^2\Big)
+\mbox{finite terms}
\Bigg).\qquad
\end{eqnarray}

\noindent
This expression is finite due to the definition of $\hat\Delta$, so
it is not necessary to add any counterterms in two- and three-loop
approximations. Thus for all renomalization schemes with linear
$\hat\Delta$ we have:

\begin{eqnarray}\label{Three_Loop_E2}
\frac{4\pi^2}{e_0^2} = \frac{\pi}{\alpha\Big(\Lambda/\mu\Big)}
- \ln\frac{\Lambda}{\mu}-b_1+O(\alpha^3).
\end{eqnarray}

\noindent
This means, that the two- and three-loop contributions to the
$\beta$-function are equal to zero and

\begin{equation}
\beta = \frac{\alpha^2}{\pi} + O(\alpha^5).
\end{equation}

\noindent
Hence the $\beta$-function is exhausted at the one-loop and agrees with
the multiplet structure of anomalies.

Note, that the sum of diagrams which do not contain insersions of
counterterms on matter lines in the considered approximation gives
the following contribution to the $\beta$-function:

\begin{equation}
\Delta\beta = \frac{\alpha^2}{\pi}\Big(1-\gamma(\alpha)\Big).
\end{equation}

\noindent
This contribution is equal to the NSVZ $\beta$-function, but the
anomalous dimension is cancelled after adding (\ref{Delta_Beta}),
and the final result is comletely defined by the one-loop.


\section{Comparison between HD regularization and DRED.}
\hspace{\parindent}
\label{Section_HD_And_DRED}

The $\beta$-function obtained in the previous section is different from
the corresponding result, found with DRED. In the two-loop approximation
the calculations of the effective action with DRED and HD were compared in
\cite{HD_And_DRED}. The difference of the results for the $\beta$-function
is shown to have originated from the different results for the sum of
diagrams with insersions of counterterms. With DRED this contribution is
0, while with HD it is given by (\ref{Delta_GammaV}). The calculations
made in this paper show, that in the three-loop approximation we have a
similar situation.

The difference of the results for the sum of diagrams with insersions
of counterterms \cite{HD_And_DRED} is caused by the mathematical
inconsistancy of DRED, pointed in \cite{Siegel2}, because this
inconsistency  leads to zero results for all anomalies. (We assume, that
there are no assumptions like $\mbox{tr}(AB)\ne\mbox{tr}(BA)$ and all
identities are valid for $n<4$.) In particular the sum of diagrams with
insersions of counterterms on the matter lines calculated with DRED is 0.

Let us discuss this in detail:

In supersymmetric theories the axial anomaly is related with the Konishi
anomaly \cite{Konishi,ClarkKonishi}. Indeed, let us consider

\begin{equation}\label{Im_Of_Konishi}
\mbox{Im}\Bigg[\bar D^2 \Big(\phi^* e^{2V}\phi
+ \tilde\phi^* e^{-2V}\tilde\phi\Big)\Bigg].
\end{equation}

\noindent
Using equations (\ref{Phi_Superfield}) and (\ref{V_Superfield}), it is
easy to see, that in components this expression will contain (among other
terms)

\begin{equation}\label{Axial_Current}
- \bar\theta\theta\,
\partial_\mu \Big(\bar\Psi \gamma^\mu\gamma_5\Psi\Big),
\end{equation}

\noindent
where the Dirac spinor $\Psi$ is defined by (\ref{Psi_Definition}).
It is well known \cite{Bertlmann}, that the conservation of the axial
current is broken by quantum corrections and in particular

\begin{equation}\label{Axial_Anomaly}
\bar\theta\theta\,
\langle \partial_\mu \Big(\bar\Psi \gamma^\mu\gamma_5\Psi\Big)\rangle
= - \bar\theta\theta\,\frac{1}{8\pi^2} F_{\mu\nu}\tilde F^{\mu\nu}.
\end{equation}

\noindent
Hence due to the supersymmetry
\footnote{Note, that our arguments can not be considered as a derivation
of the Konishi anomaly, because (\ref{Axial_Current}) does not contain
all terms of (\ref{Im_Of_Konishi}), proportional to $\bar\theta\theta$.
A strict derivation of the Konishi anomaly can be found in
\cite{Konishi,ClarkKonishi}. Our goal is only to remind of the relation
between the axial anomaly and the Konishi anomaly.}

\begin{equation}\label{Im_Of_Anomaly}
\mbox{Im}\,\Big\langle \bar D^2 \Big(\phi^* e^{2V}\phi
+ \tilde\phi^* e^{-2V}\tilde\phi\Big)\Big\rangle
= \frac{1}{2\pi^2} \mbox{Im}\Big(W_a C^{ab} W_b\Big).
\end{equation}

\noindent
By performing supersymmetry transformations it is easy to see, that if
an imaginary part of a chiral superfield is equal to 0, then this
superfield is a real constant. Therefore, from (\ref{Im_Of_Anomaly}) we
obtain, that

\begin{equation}\label{Full_Anomaly}
\Big\langle \bar D^2 \Big(\phi^* e^{2V}\phi
+ \tilde\phi^* e^{-2V}\tilde\phi\Big)\Big\rangle
= \frac{1}{2\pi^2} W_a C^{ab} W_b + \mbox{const}.
\end{equation}

\noindent
Applying

\begin{equation}
-\frac{1}{2}\int d^4x\, D^2 = \int d^4x\,d^2\theta
\end{equation}

\noindent
to (\ref{Full_Anomaly}) and taking a real part of the result, we obtain,
that

\begin{equation}\label{Konishi_Anomaly}
\Big\langle \frac{1}{4}\int d^4x\,d^4\theta\,\Big(\phi^* e^{2V}\phi
+ \tilde\phi^* e^{-2V}\tilde\phi\Big)\Big\rangle
= - \frac{1}{16\pi^2} \mbox{Re}\int d^4x\,d^2\theta\,W_a C^{ab} W_b.
\end{equation}

Because DRED requires, that the space-time dimension $n$ should be less
than 4 \cite{Siegel}, it is possible to choose $\gamma_5$ anticommuting
with all $\gamma$-matrices. Then the chiral symmetry is not broken in the
regularized theory due to the mathematical inconsistency of DRED. As a
consequence axial anomaly appears to be 0, while the supersymmetry is not
broken. Therefore instead of (\ref{Konishi_Anomaly}) we obtain

\begin{equation}\label{Incorrect_Konishi}
\Big\langle \int d^4x\,d^4\theta\,\Big(\phi^* e^{2V}\phi
+ \tilde\phi^* e^{-2V}\tilde\phi\Big)\Big\rangle = 0.
\end{equation}

\noindent
In DREG such problem can be solved by using $\gamma_5$ with the
following properties:

\begin{equation}
\{\gamma_5,\gamma_\mu\}=0,\quad \mu=0,\ldots,3;\qquad
[\gamma_5,\gamma_\mu]=0,\quad \mu>3.
\end{equation}

\noindent
Then the chiral symmetry is broken in the regularized theory, and
axial anomaly is calculated correctly \cite{tHV}. Nevertheless,
DREG breaks the supersymmetry and is not well-suited for supersymmetric
theories.

As a consequence of (\ref{Konishi_Anomaly}) we obtain \cite{HD_And_DRED}
the identity

\begin{eqnarray}\label{Generalization_Of_Konishi_Anomaly}
&& \Big\langle \exp\Bigg(i (Z-1)\,
\frac{1}{4} \int d^4x\,d^4\theta\,\Big(\phi^* e^{2V}\phi
+ \tilde\phi^* e^{-2V}\tilde\phi\Big)\Bigg)\Big\rangle
=\nonumber\\
&& = \exp\Bigg(-i\ln Z
\frac{1}{16\pi^2} \mbox{Re}\int d^4x\,d^2\theta\,W_a C^{ab} W_b
+\mbox{finite terms}
\Bigg),
\qquad
\end{eqnarray}

\noindent
whose l.h.s. is a sum of all diagrams with insersions of counterterms
on lines of the matter superfield. The corresponding result obtained
with DRED, which follows from (\ref{Incorrect_Konishi}), is written as

\begin{equation}
\Big\langle \exp\Bigg(i (Z-1)\,
\frac{1}{4} \int d^4x\,d^4\theta\,\Big(\phi^* e^{2V}\phi
+ \tilde\phi^* e^{-2V}\tilde\phi\Big)\Bigg)\Big\rangle = 1.
\end{equation}

\noindent
Then the sum of all diagrams with insersions of counterterms is 0, that
contradicts the result for Konishi anomaly. Thus the mathematical
inconsistency of DRED gives the result for the $\beta$-function which
differs from the corresponding result obtained with HD and leads to the
anomaly puzzle.


\section{Solution of the anomaly puzzle}
\hspace{\parindent}
\label{Section_Anomaly_Puzzle}

In order to investigate the anomaly puzzle it is convenient to use
the higher derivative regularization, because it is applicable for
the calculation of anomalies.

From the calculation described in section \ref{Section_Three_Loop}
we see, that the $\beta$-function, defined by (\ref{Beta_Gamma_Definition})
is exausted at the one-loop, while the $\beta$-function defined by
(\ref{GL_Function}) has corrections from all orders. This seems to
contradict equation (\ref{GL_And_Beta}). However, actually there is
no contradiction, because the generating functional (\ref{Modified_Z})
depends on $\mu$. Really, due to the rescaling anomaly
(\ref{Generalization_Of_Konishi_Anomaly}) it is impossible to remove
$\mu$-dependence by the transformation $\phi\to Z^{-1/2}\phi$, because
the anomalous contribution contains $\mu$-dependent $\ln Z$. Therefore,
the $\beta$-functions (\ref{Beta_Gamma_Definition}) and (\ref{GL_Function})
are different. (See the derivation of equation (\ref{GL_And_Beta}).) The
first function is proportional to the trace anomaly and due to the
multiplet structure of anomalies is exausted by the first loop, while
the second one has corrections from all orders.

Nevertheless, it is desirable to avoid $\mu$-dependence of the generating
functional. It can be made by two different ways. We can assume, that the
bare coupling constant $e_0$ depends on $\mu$. In this case $\beta$-function
(\ref{Beta_Gamma_Definition}) will have corrections from all orders, but
it will not be proportional to the trace anomaly. Another possibility is
to use canonical normalization for the matter superfields and to define
the generating functional by

\begin{eqnarray}\label{Canonical_Z}
&& Z = \int DV\,D\phi\,D\tilde \phi\,
\prod\limits_i \Big(\det{}' PV(V,M_i)\Big)^{c_i}
\exp\Bigg\{i\Bigg[\frac{1}{4 e^2} \int d^4x\,d^4\theta\, V\partial^2
\Big(1+ \frac{\partial^{2n}}{\Lambda^{2n}}\Big) V
-\nonumber\\
&& - \frac{1}{4 e^2} \Big(Z_3(\Lambda/\mu)-1\Big) \int d^4x\,d^4\theta\,
V \Pi_{1/2}\partial^2
\Big(1+ \frac{\partial^{2n}}{\Lambda^{2n}}\Big) V
+\nonumber\\
&& \qquad\qquad\qquad\qquad\qquad\qquad\qquad\qquad
+ \frac{1}{4} \int d^4x\,d^4\theta\,
\Big(\phi^* e^{2V}\phi
+ \tilde\phi^* e^{-2V}\tilde\phi \Big)
+\nonumber\\
&&
+ \int d^4x\,d^4\theta\,J V
+ \int d^4x\,d^2\theta\, \Big(j\,\phi + \tilde j\,\tilde\phi \Big)
+ \int d^4x\,d^2\bar\theta\,
\Big(j^*\phi^* + \tilde j^* \tilde\phi^* \Big)\Bigg]\Bigg\},
\end{eqnarray}

\noindent
where

\begin{eqnarray}\label{Canonical_PV_Determinants}
&& \Big(\det{}' PV(V,M)\Big)^{-1} \equiv \int D\Phi\,D\tilde \Phi\,
\exp\Bigg\{i\Bigg[ \frac{1}{4} \int d^4x\,d^4\theta\,
\Big(\Phi^* e^{2V}\Phi
+\qquad\nonumber\\
&& \qquad\qquad\qquad
+ \tilde\Phi^* e^{-2V}\tilde\Phi \Big)
+ \frac{1}{2}\int d^4x\,d^2\theta\, M \tilde\Phi \Phi
+ \frac{1}{2}\int d^4x\,d^2\bar\theta\, M \tilde\Phi^* \Phi^*
\Bigg]\Bigg\}.\qquad
\end{eqnarray}

\noindent
Then the $\beta$-function (\ref{Beta_Gamma_Definition}) will be
proportional to the trace anomaly, but Adler-Bardeen theorem is not
valid in this case.


\section{Conclusion}
\hspace{\parindent}
\label{Section_Conclusion}

In this paper we calculated the three-loop $\beta$-function for $N=1$ SUSY
QED regularized by higher derivatives. Using the standard definition of
the generating functional we found, that two- and three-loop contributions
to the $\beta$-function (\ref{Beta_Gamma_Definition}) were 0 for a large
number of subtraction schemes. In this case the sum of diagrams without
insersions of counterterms on matter lines is exactly equal to the terms
of the corresponding order in the expansion of the NSVZ $\beta$-function.
However two- and three-loop contributions are exactly cancelled by diagrams
containing insersions of counterterms.

The result for $\beta$-function (\ref{Beta_Gamma_Definition}) obtained
with HD regularization differs from the corresponding result obtained with
DRED, because DRED is not mathematically consistent and does not permit to
calculate anomalies \cite{HD_And_DRED} (if there are no additional
assumtions like $\mbox{tr}(AB)\ne \mbox{tr}(BA)$ e.t.c.). In particular,
the Konishi anomaly, which contributes to $\beta$-function
(\ref{Beta_Gamma_Definition}), calculated with DRED is 0. This in turn
leads to the anomaly puzzle. HD regularization enables us to find an
anomalous contribution of diagrams with insersions of the counterterms,
which was calculated in \cite{HD_And_DRED} exactly to all orders and is
equal to

\begin{equation}
\frac{\alpha^2}{\pi} \gamma(\alpha).
\end{equation}

\noindent
The calculations done in this paper confirm this result in the
three-loop approximation.

The result for $\beta$-function (\ref{Beta_Gamma_Definition}), obtained
with the generating functional (\ref{Modified_Z}), is consistent with a
multiplet structure of the anomalies: Since in supersymmetric theories
the axial and the trace of the energy-momentum tensor anomalies are
members of a supersymmetric multiplet, the $\beta$-function
(\ref{Beta_Gamma_Definition}) should be exhausted by the first loop. In
particular, if the theory is regularized by HD the Adler-Bardeen theorem
does not conflict with supersymmetry, while for theories, regularized by
DRED, such a contradiction seems to take place \cite{Kazakov}.

However, the generating functional (\ref{Modified_Z}) depends on $\mu$
due to the rescaling anomaly. As a consequence, the $\beta$-function
(\ref{GL_Function}) is different from the one defined by equation
(\ref{Beta_Gamma_Definition}). If we would like to define a
$\mu$-independent generating functional, then either Adler-Bardeen
theorem is not valid or the trace anomaly is not proportional to the
$\beta$-function. Therefore, if the generating functional does not
depend on the normalization point, then the arguments based on the
multiplet structure of anomalies can not be used. In our opinion this
solves the anomaly puzzle in the considered model.

One of the possible ways to define a $\mu$-independent generating
functional is the using of the canonical normalization
(\ref{Canonical_Normalization}). Then there are no diagrams with
insersions of counterterms and the $\beta$-function is equal to
the NSVZ expression. It is important to note, that unlike DRED the HD
regularization does not require to tune the subtraction scheme. The
NSVZ expression (at least in the three-loop approximation) for
$\beta$-function (\ref{Beta_Gamma_Definition}) is authomatically
obtained with HD regularization if an operator, constructing a counterterm
for a given function, is linear.

It is necessary to note, that so far we considered only the Abelian case.
For the supersymmetric Yang-Mills theory the using of higher covariant
derivative regularization \cite{West_Paper} leads to very involved
calculations, because in this case Feynman rules become much more
complicated. In this case the using of usual derivatives can simplify
the calculations considerably. However, such regularization breaks the
gauge invariance. Nevertheless, even in the case of noninvariant
regularization it is possible to obtain the gauge invariant renormalized
effective action by a special choice of subtraction scheme
\cite{Slavnov1,Slavnov2}. For Abelian supersymmetric theories such scheme
was proposed in \cite{SlavnovStepan1}. Construction of the invariant
renormalization procedure for supersymmetric non-Abelian models is in
progress.


\vspace{1cm}

\noindent
{\Large{\bf Acknowledgments}}

\bigskip

We would like to express our gratitude to D.I.Kazakov, V.A.Novikov,
M.Perez-Victoria, P.I.Pronin, A.A.Slavnov and to our colleagues from
Institute of Theoretical and Experimental Physics, Joint Institute of
Nuclear Research and Steklov Mathematical Institute for valuable
discussions.


\vspace{1cm}

\noindent
{\Large\bf Appendix.}


\appendix


\section{Results for Feynman diagrams.}
\hspace{\parindent}
\label{Appendix_Diagrams}

Having calculated Feynman diagrams, presented in Fig.
\ref{Figure_Beta_Diagrams_1Loop} -- \ref{Figure_Beta_Diagrams_XX},
we obtained the following expressions in the Minkowski space:
\footnote{For simplicity we omit $+i0$ in propagators.}

1. One-loop diagrams, presented in Fig.\ref{Figure_Beta_Diagrams_1Loop}:

\begin{equation}
f_{1-loop} = -\frac{i}{2}
\int \frac{d^4k}{(2\pi)^4} \Bigg(\frac{1}{k^2 (k+p)^2}
- \sum\limits_i c_i\frac{1}{(k^2-M_i^2)\Big((k+p)^2-M_i^2\Big)}\Bigg).
\end{equation}

2. Two-loop diagrams, presented in Fig.\ref{Figure_Beta_Diagrams_2Loop},
and three-loop diagrams, presented in Fig. \ref{Figure_Beta_Diagrams_Oo},
with the external loop of $\phi$ and $\tilde\phi$:

\begin{eqnarray}
&& f_{Oo}
= - e^2 \int \frac{d^4k}{(2\pi)^4}\frac{d^4q}{(2\pi)^4}
\frac{(k+p+q)^2+q^2-k^2-p^2}{k^2\Big(1 + (-1)^n k^{2n}/\Lambda^{2n}\Big)^2
(k+q)^2 (k+p+q)^2 q^2 (q+p)^2}
\times\ \nonumber\\
&&\times
\Bigg[
\Big(1+(-1)^n k^{2n}/\Lambda^{2n}\Big)
\Bigg(1+\frac{e^2}{4\pi^2}\Big(\ln\frac{\Lambda}{\mu}+b_1\Big)\Bigg)
+ 2i e^2
\Bigg(\int \frac{d^4t}{(2\pi)^4}\,\frac{1}{t^2 (k+t)^2}
-\nonumber\\
&& - \sum\limits_i c_i \int \frac{d^4t}{(2\pi)^4}\,
\frac{1}{\Big(t^2-M_i^2\Big) \Big((k+t)^2-M_i^2\Big)}\Bigg)\Bigg].\qquad
\end{eqnarray}

3. The same diagrams, with the extenal loop of Pauli-Villars fields:

\begin{eqnarray}
&& f_{Oo_{PV}}= e^2 \sum\limits_j c_j\,
\int \frac{d^4k}{(2\pi)^4}\frac{d^4q}{(2\pi)^4}\,
\frac{1}{k^2 \Big(1 + (-1)^n k^{2n}/\Lambda^{2n}\Big)^2}
\times\nonumber\\
&& \times
\Bigg[\frac{(k+p+q)^2+q^2-k^2-p^2}{
\Big((k+q)^2-M_j^2\Big) \Big((k+p+q)^2-M_j^2\Big) \Big(q^2-M_j^2\Big)
\Big((q+p)^2-M_j^2\Big)}
+\nonumber\\
&& \qquad\qquad\qquad\qquad\qquad\qquad
+ \frac{4 M_j^2}{\Big((k+q)^2-M_j^2\Big) \Big(q^2-M_j^2\Big)^2
\Big((q+p)^2-M_j^2\Big)} \Bigg]
\times\nonumber\\
&&\times
\Bigg[
\Big(1+(-1)^n k^{2n}/\Lambda^{2n}\Big)
\Bigg(1+\frac{e^2}{4\pi^2}\Big(\ln\frac{\Lambda}{\mu}+b_1\Big)\Bigg)
+ 2i e^2
\Bigg(\int \frac{d^4t}{(2\pi)^4}\,\frac{1}{t^2 (k+t)^2}
-\nonumber\\
&& - \sum\limits_i c_i \int \frac{d^4t}{(2\pi)^4}\,
\frac{1}{\Big(t^2-M_i^2\Big) \Big((k+t)^2-M_i^2\Big)}\Bigg)\Bigg].\qquad
\end{eqnarray}

4. Diagrams with two loops of matter superfields, presented in Fig.
\ref{Figure_Beta_Diagrams_OO}:

\begin{equation}
f_{OO}=0.
\end{equation}

5. Diagrams with a single loop of $\phi$ and $\tilde\phi$, presented
in Fig. \ref{Figure_Beta_Diagrams_O}

\begin{eqnarray}\label{O_Diagrams}
&& f_{O} = -i\int \frac{d^4k}{(2\pi)^4}\,\frac{d^4l}{(2\pi)^4}\,
\frac{d^4q}{(2\pi)^4}\times\nonumber\\
&& \times
\frac{e^4}{k^2 \Big(1+(-1)^n k^{2n}/\Lambda^{2n}\Big)
\,l^2\Big(1+(-1)^n l^{2n}/\Lambda^{2n}\Big)\,q^2 (q+p)^2 (q+k+l)^2}
\times\nonumber\\
&& \times
\Bigg\{
\frac{2 k^2 \Big((2q+k+l)^2+(2q+k+l+p)^2-p^2\Big)}{(q+k)^2
(q+l)^2 (q+k+p)^2}
- \frac{4 p^2}{(q+k)^2 (q+k+p)^2}
-\nonumber\\
&& - \frac{4(q+p)^2}{(q+k)^2 (q+k+p)^2}
+ \frac{4(q+p)^2}{(q+k+p)^2 (q+k+l+p)^2}
+ \frac{4}{(k+q)^2}
-\nonumber\\
&& - \frac{2p^4}{(q+k)^2 (q+k+p)^2 (q+k+l+p)^2}
- \frac{4 k^2}{(q+k)^2 (q+k+p)^2}
-\\
&& - \frac{2 k^2 p^2}{(q+k)^2 (q+k+p)^2 (q+k+l+p)^2}
- \frac{2 (k+l)^2}{(q+k+l+p)^2 (q+k)^2}
+\nonumber\\
&& + \frac{(k+l)^2 p^2}{(q+k)^2 (q+k+p)^2 (q+k+l+p)^2}
- \frac{4(2q+k+l+p)^2}{(q+k+p)^2 (q+l)^2}
+\nonumber\\
&& + \frac{2 (k+l)^2 (2q+k+l+p)^2}{(q+k+l+p)^2 (q+l)^2 (q+k+p)^2}
+ \frac{2p^2 (2q+k+l+p)^2}{(q+k+l+p)^2 (q+l)^2 (q+k+p)^2}
\Bigg\}.\nonumber
\end{eqnarray}

6. The same diagrams with the Pauli-Villars loop

\begin{eqnarray}
&& f_{O_{PV}} = i\sum\limits_i c_i
\int \frac{d^4k}{(2\pi)^4}\,\frac{d^4l}{(2\pi)^4}\,
\frac{d^4q}{(2\pi)^4}\,
\frac{e^4}{k^2 \Big(1+(-1)^n k^{2n}/\Lambda^{2n}\Big)
\,l^2\Big(1+(-1)^n l^{2n}/\Lambda^{2n}\Big)}
\times\nonumber\\
&& \times
\frac{1}{\Big(q^2-M_i^2\Big)
\Big((q+p)^2-M_i^2\Big) \Big((q+k)^2-M_i^2\Big) \Big((q+k+l)^2-M_i^2\Big)}
\times\nonumber\\
&& \times
\Bigg\{
\frac{2}{\Big((q+l)^2-M_i^2\Big)\Big((q+k+p)^2-M_i^2\Big)}
\Bigg[-2(q+k+l)^2 \Big((q+k+p)^2
+\nonumber\\
&& +(q+k)^2 - p^2\Big)
+ k^2 \Big((2q+k+l)^2 + (2q+k+l+p)^2 - p^2\Big) - 8 M_i^4
-\vphantom{\Bigg(}\nonumber\\
&& - 2 M_i^2 (2q+k+l)^2
- 2 M_i^2 (2q+k+l+p)^2
+ 2 M_i^2\Big(4(q+k+p)^2
+\vphantom{\Bigg(}\nonumber\\
&&
+ 4(q+k)^2-k^2-p^2\Big)
+ 2 M_i^2 (q+l)_\mu (2q+2k+2l-2p)_\mu \Bigg]
+\nonumber\\
&& +\frac{1}{\Big((q+k+p)^2-M_i^2\Big)\Big((q+k+l+p)^2-M_i^2\Big)}
\Bigg[-2q^2 (q+k+l+p)^2
-\nonumber\\
&&
- 4 (q+p)^2 (q+k+l+p)^2 - 8 (q+k+p)^2 (q+p)^2
-4 q^2 (q+k+p)^2
+\vphantom{\Bigg(}\nonumber\\
&& + 2p^2 \Big(6M_i^2-p^2\Big) +2l^2 \Big(2(q+p)^2-p^2\Big)
+ (k+l)^2 \Big(2(q+k+p)^2-p^2
+\vphantom{\Bigg(}\nonumber\\
&& +2M_i^2\Big)
- 6 M_i^4 + 4M_i^2\Big(2(q+k+p)^2+2(q+p)^2-k^2-p^2\Big)
\Bigg]
+\nonumber\\
&& +\frac{2}{\Big((q+l+p)^2-M_i^2\Big)\Big((q+k+l+p)^2-M_i^2\Big)}
\Bigg[-6M_i^4+(2q+k+l+p)^2
\times\nonumber\\
&& \times
\Big(-2(q+k+l+p)^2+(k+l)^2+p^2\Big)
+4M_i^2\Big(3(q+k+l+p)^2-p^2\Big)
-\vphantom{\Bigg(}\nonumber\\
&&
-3M_i^2(k+l)^2-2M_i^2(k+q)^2+2M_i^2(l+p)^2\Bigg]
-\frac{16M_i^2}{\Big((q+l+p)^2-M_i^2\Big)}
+\nonumber\\
&& +\frac{16}{\Big((q+k)^2-M_i^2\Big)\Big((q+k+p)^2-M_i^2\Big)}\Bigg[
\Big((q+k)^2-3M_i^2\Big)\Big((q+k+p)^2
+\nonumber\\
&& +(q+k)^2+q^2+(q+p)^2-k^2\Big)
+2M_i^2p^2+5M_i^4-M_i^2(q+k)^2 \Bigg]
+\\
&& +\frac{2\Big(2(q+k+l)^2-(k+l)^2\Big)\Big((q+k)^2-M_i^2\Big)
}{\Big((q+k+p)^2-M_i^2\Big)\Big((q+k+l+p)^2-M_i^2\Big)}
-\frac{4M_i^2}{\Big((q+l)^2-M_i^2\Big)}
+\vphantom{\Bigg(}\nonumber\\
&& +\frac{2\Big(-q^2(q+p)^2+6M_i^2q^2-9M_i^4\Big)\Big((q+k+l)^2-M_i^2\Big)
}{(q^2-M_i^2)\Big((q+p)^2-M_i^2\Big)\Big((q+l+p)^2-M_i^2\Big)}
+ \frac{16M_i^2}{q^2-M_i^2}
+\vphantom{\Bigg(}\nonumber\\
&& +\frac{4\Big(-q^4+3M_i^2 q^2-6M_i^4\Big)\Big((q+k+l)^2-M_i^2\Big)
}{(q^2-M_i^2)^2\Big((q+l)^2-M_i^2\Big)}
-\frac{8M_i^2 (2q+k+l)^2-16M_i^4}{(q^2-M_i^2)\Big((q+l)^2-M_i^2\Big)}
+\vphantom{\Bigg(}\nonumber\\
&& +\frac{4}{(q^2-M_i^2)\Big((q+k)^2-M_i^2\Big)}\Bigg[-(q+k)^2 q^2
+ 3M_i^2 q^2 - M_i^2 (q+k)^2 - 5M_i^4\Bigg]\Bigg\}.\nonumber
\end{eqnarray}

7. Diagrams containing one insersion of counterterms, presented in Fig.
\ref{Figure_Beta_Diagrams_X}:

\begin{eqnarray}\label{X_Diagrams}
&& f_{X} =
- 2i \Bigg(
\frac{\alpha}{\pi}\Big(\ln\frac{\Lambda}{\mu}+g_1\Big)
+ \frac{\alpha^2}{\pi^2}\Big(\ln^2\frac{\Lambda}{\mu}
+ g_1 \ln\frac{\Lambda}{\mu}\Big)
- \gamma_2\,\alpha^2 \Big(\ln\frac{\Lambda}{\mu}+g_2\Big)\Bigg)
\times\quad\nonumber\\
&& \times
\sum\limits_i c_i
\int \frac{d^4k}{(2\pi)^4}\,\frac{M_i^2}{(k^2-M_i^2)^2
\Big((k+p)^2-M_i^2\Big)}.\qquad
\end{eqnarray}

8. Diagrams containing two insersions of counterterms, presented in Fig.
\ref{Figure_Beta_Diagrams_XX}:

\begin{eqnarray}\label{XX_Diagrams}
&& f_{XX} = \frac{i\alpha^2}{\pi^2}\Big(\ln\frac{\Lambda}{\mu}+g_1\Big)^2
\times\nonumber\\
&& \qquad
\times
\sum\limits_i c_i M_i^2
\int \frac{d^4k}{(2\pi)^4} \frac{3 k^2 (k+p)^2 - 6 M_i^4 + 3 k^4
- 5 k^2 M_i^2 + 5 (k+p)^2 M_i^2}{
2 \Big(k^2-M_i^2\Big)^3 \Big((k+p)^2-M_i^2\Big)^2}.\qquad
\end{eqnarray}

9. Two-loop diagrams containing insersion of counterterms, presented
in Fig. \ref{Figure_Beta_Diagrams_X2}:

\begin{eqnarray}
&& f_{X2} = e^2 \sum\limits_i c_i
\frac{\alpha}{\pi}\Big(\ln\frac{\Lambda}{\mu}+g_1\Big)
\int \frac{d^4k}{(2\pi)^4}\frac{d^4q}{(2\pi)^4}\,
\frac{4 M_i^2}{k^2 \Big(1+(-1)^n k^{2n}/\Lambda^{2n}\Big)}
\times\nonumber\\
&& \times
\Bigg\{
\frac{(k+q+p)^2+(k+q)^2+(q+p)^2+q^2-2k^2-2p^2}{
\Big((k+q)^2-M_i^2\Big) \Big((k+q+p)^2-M_i^2\Big)
\Big((q+p)^2-M_i^2\Big)\Big(q^2-M_i^2\Big)^2}
+\nonumber\\
&& + \frac{2(k+q)^2}{\Big((k+q)^2-M_i^2\Big)^2
\Big((q+p)^2-M_i^2\Big) \Big(q^2-M_i^2\Big)^2}
+\nonumber\\
&& \qquad\qquad\qquad\qquad\qquad
+ \frac{4(q+p)^2 M_i^2+2q^2 M_i^2-6M_i^4}{\Big((k+q)^2-M_i^2\Big)
\Big((q+p)^2-M_i^2\Big)^2 \Big(q^2-M_i^2\Big)^3}
\Bigg\}.\qquad
\end{eqnarray}


\section{Three-loop contributions to the $\beta$-function.}
\hspace{\parindent}
\label{Appendix_Relation}

In order to find the three-loop $\beta$-function, it is necessary to
calculate the integrals, presented in Appendix \ref{Appendix_Diagrams}.

1. Performing the Wick rotation and using the standard technique, it is
easy to see, that

\begin{eqnarray}
&& f_{1-loop} = \frac{1}{2}
\int \frac{d^4k}{(2\pi)^4} \Bigg(\frac{1}{k^2 (k+p)^2}
- \sum\limits_i c_i\frac{1}{(k^2+M_i^2)\Big((k+p)^2+M_i^2\Big)}\Bigg)
=\nonumber\\
&& = \frac{1}{16\pi^2} \sum\limits_i c_i
\Bigg(\ln\frac{M_i}{p} + \sqrt{1+\frac{4M_i^2}{p^2}}
\mbox{arctanh}\sqrt{\frac{p^2}{4M_i^2 + p^2}}\Bigg)
= \frac{1}{16\pi^2} \ln \frac{\Lambda}{p} + O(1).\qquad
\end{eqnarray}

\noindent
(We assume, that $M_i=a_i\Lambda$, where $a_i$ are some constants.)

2. Next we analyze graphs containing insersions of counterterms on
matter lines: Integrals in $f_X$, $f_{XX}$ and $f_{X2}$ are functions
of $p/\Lambda$ finite at $\Lambda\to\infty$. Therefore, the divergent
part of the effective action is defined by their values at $p=0$.
In this limit expressions (\ref{X_Diagrams}) and (\ref{XX_Diagrams})
in Euclidean space can be written as

\begin{eqnarray}\label{FX}
&& f_{X} = -2
\Bigg(\frac{\alpha}{\pi}\Big(\ln\frac{\Lambda}{\mu}+g_1\Big)
+ \frac{\alpha^2}{\pi^2}\Big(\ln^2\frac{\Lambda}{\mu}
+ g_1 \ln\frac{\Lambda}{\mu}\Big)
- \gamma_2\,\alpha^2 \Big(\ln\frac{\Lambda}{\mu}+g_2\Big) \Bigg)
\times\nonumber\\
&& \qquad\qquad\qquad\qquad\qquad\qquad\qquad\qquad\qquad\quad
\times
\sum\limits_i c_i \int \frac{d^4k}{(2\pi)^4}\,\frac{M_i^2}{(k^2+M_i^2)^3}
=\qquad\nonumber\\
&& = -\frac{1}{16\pi^2}
\Bigg(
\frac{\alpha}{\pi}\Big(\ln\frac{\Lambda}{\mu}+g_1\Big)
+ \frac{\alpha^2}{\pi^2}\Big(\ln^2\frac{\Lambda}{\mu}
+ g_1 \ln\frac{\Lambda}{\mu}\Big)
- \gamma_2\,\alpha^2 \Big(\ln\frac{\Lambda}{\mu}+g_2\Big)
\Bigg);\qquad\\
&&\vphantom{\Big(}\nonumber\\
\label{FXX}
&& f_{XX} = \frac{e^4}{16\pi^4}\Big(\ln\frac{\Lambda}{\mu}+g_1\Big)^2
\sum\limits_i c_i
\int \frac{d^4k}{(2\pi)^4}
\frac{3 (k^2 - M_i^2) M_i^2}{\Big(k^2+M_i^2\Big)^4}
= \frac{1}{16\pi^2}\,\frac{\alpha^2}{2\pi^2}
\Big(\ln\frac{\Lambda}{\mu}+g_1\Big)^2,\nonumber\\
\end{eqnarray}

\noindent
where we take into account, that $\sum c_i = 1$. In order to calculate
$f_{X2}$ we note, that at $p\to 0$

\begin{eqnarray}\label{FX2_At_P=0}
&& f_{X2} = -\Bigg(\frac{\alpha}{\pi}\ln\frac{\Lambda}{\mu}+g_1\Bigg)\,
\sum\limits_i c_i\, M_i \frac{d}{dM_i} \int\frac{d^4k}{(2\pi)^4}
\frac{d^4q}{(2\pi)^4}\,\frac{e^2}{k^2\Big(1+k^{2n}/\Lambda^{2n}\Big)}
\times\nonumber\\
&& \qquad\qquad
\times
\Bigg\{\frac{(k+q)^2+q^2-k^2}{\Big((k+q)^2+M_i^2\Big)^2
\Big(q^2+M_i^2\Big)^2} - \frac{4 M_i^2}{\Big((k+q)^2+M_i^2\Big)
\Big(q^2+M_i^2\Big)^3} \Bigg\}.\qquad
\end{eqnarray}

\noindent
In appendix \ref{Appendix_Integrals} we prove, that this integral is
equal to 0, and therefore

\begin{equation}\label{FX2}
f_{X2} = 0.
\end{equation}

\noindent
Expressions (\ref{FX}), (\ref{FXX}) and (\ref{FX2}) agree with the exact
result for the sum of diagrams with insersions of counterterms, obtained
in \cite{HD_And_DRED}:

\begin{equation}\label{Exact_Result}
- \ln Z\,\frac{1}{16\pi^2}\,
\mbox{Re}\int d^4x\,d^2\theta\,W_a C^{ab} W_b +\mbox{finite terms}.
\end{equation}

\noindent
And indeed, for the considered theory $Z$ is given by (\ref{Constant_Z})
and the terms of the considered order in $\alpha$ in (\ref{Exact_Result})
are

\begin{eqnarray}
&& - \frac{1}{16\pi^2}\,\Bigg(
\frac{\alpha}{\pi}\Big(\ln\frac{\Lambda}{\mu}+g_1\Big)
+ \frac{\alpha^2}{\pi^2}\Big(\ln^2\frac{\Lambda}{\mu}
+ g_1 \ln\frac{\Lambda}{\mu}\Big)
- \gamma_2\,\alpha^2 \Big(\ln\frac{\Lambda}{\mu}+g_2\Big)
-\nonumber\\
&& \qquad\qquad\qquad\qquad
- \frac{\alpha^2}{2\pi^2}\,\Big(\ln \frac{\Lambda}{\mu}+g_1\Big)^2
+ O(\alpha^3)\Bigg)\,
\mbox{Re}\int d^4x\,d^2\theta\,W_a C^{ab} W_b
=\quad\nonumber\\
&& = \Big(f_X+f_{XX}+f_{X2}+O(\alpha^3)\Big)\,\mbox{Re}
\int d^4x\,d^2\theta\,W_a C^{ab} W_b.
\end{eqnarray}

\noindent
It means, that the result for the sum of Feynman diagrams agrees
with the exact result (\ref{Generalization_Of_Konishi_Anomaly}).

\bigskip

3. In Euclidean space $f_{Oo}$ is given by

\begin{eqnarray}
&& f_{Oo}
= e^2 \int \frac{d^4k}{(2\pi)^4}\frac{d^4q}{(2\pi)^4}
\frac{(k+p+q)^2+q^2-k^2-p^2}{k^2\Big(1 + k^{2n}/\Lambda^{2n}\Big)^2
(k+q)^2 (k+p+q)^2 q^2 (q+p)^2}
\times\qquad\nonumber\\
&&\times
\Bigg[\Big(1+k^{2n}/\Lambda^{2n}\Big)
\Bigg(1+\frac{e^2}{4\pi^2}\Big(\ln\frac{\Lambda}{\mu}+b_1\Big)\Bigg)
- 2 e^2 \Bigg(\int \frac{d^4t}{(2\pi)^4}\,\frac{1}{t^2 (k+t)^2}
-\nonumber\\
&& - \sum\limits_i c_i \int \frac{d^4t}{(2\pi)^4}\,
\frac{1}{\Big(t^2+M_i^2\Big) \Big((k+t)^2+M_i^2\Big)}
\Bigg)\Bigg].
\end{eqnarray}

\noindent
In order to prove that this expression contains only the first degree of
$\ln\Lambda$, it is necessary to verify the existance of the limit

\begin{eqnarray}\label{Oo_Limit}
&& \Lambda\frac{d f_{Oo}}{d\Lambda}\Bigg|_{p=0}
= 2e^2 \,\Lambda\frac{d}{d\Lambda}
\int \frac{d^4k}{(2\pi)^4}\frac{d^4q}{(2\pi)^4}
\frac{q_\mu(q_\mu+k_\mu)}{k^2\Big(1 + k^{2n}/\Lambda^{2n}\Big)^2
(k+q)^4 q^4}
\times\nonumber\\
&&\times
\Bigg[\Big(1+k^{2n}/\Lambda^{2n}\Big)
\Bigg(1+\frac{e^2}{4\pi^2}\Big(\ln\frac{\Lambda}{\mu}+b_1\Big)\Bigg)
- 2 e^2 \Bigg(\int \frac{d^4t}{(2\pi)^4}\,\frac{1}{t^2 (k+t)^2}
-\nonumber\\
&& - \sum\limits_i c_i \int \frac{d^4t}{(2\pi)^4}\,
\frac{1}{\Big(t^2+M_i^2\Big) \Big((k+t)^2+M_i^2\Big)}
\Bigg)\Bigg].
\end{eqnarray}

\noindent
Taking into account, that

\begin{equation}
\int d^4q\,\frac{q_\mu (k_\mu+q_\mu)}{(k+q)^4 q^4} = \frac{\pi^2}{k^2}
\end{equation}

\noindent
(this identity is derived in Appendix \ref{Appendix_Integrals}),
(\ref{Oo_Limit}) can be written as

\begin{eqnarray}\label{Beta_Oo}
&& \Lambda\frac{d f_{Oo}}{d\Lambda}\Bigg|_{p=0}
=\nonumber\\
&& = \frac{e^2}{8\pi^2}\,\Lambda\frac{d}{d\Lambda}
\int \frac{d^4k}{(2\pi)^4}
\frac{1}{k^4\Big(1 + k^{2n}/\Lambda^{2n}\Big)^2}
\Bigg[\Big(1+k^{2n}/\Lambda^{2n}\Big)
\Bigg(1+\frac{e^2}{4\pi^2}\Big(\ln\frac{\Lambda}{\mu}+b_1\Big)\Bigg)
-\nonumber\\
&& - 2 e^2 \Bigg(\int \frac{d^4t}{(2\pi)^4}\,\frac{1}{t^2 (k+t)^2}
- \sum\limits_i c_i \int \frac{d^4t}{(2\pi)^4}\,
\frac{1}{\Big(t^2+M_i^2\Big) \Big((k+t)^2+M_i^2\Big)}
\Bigg)\Bigg].
\end{eqnarray}

\begin{figure}[h]
\epsfxsize15.0truecm\epsfbox{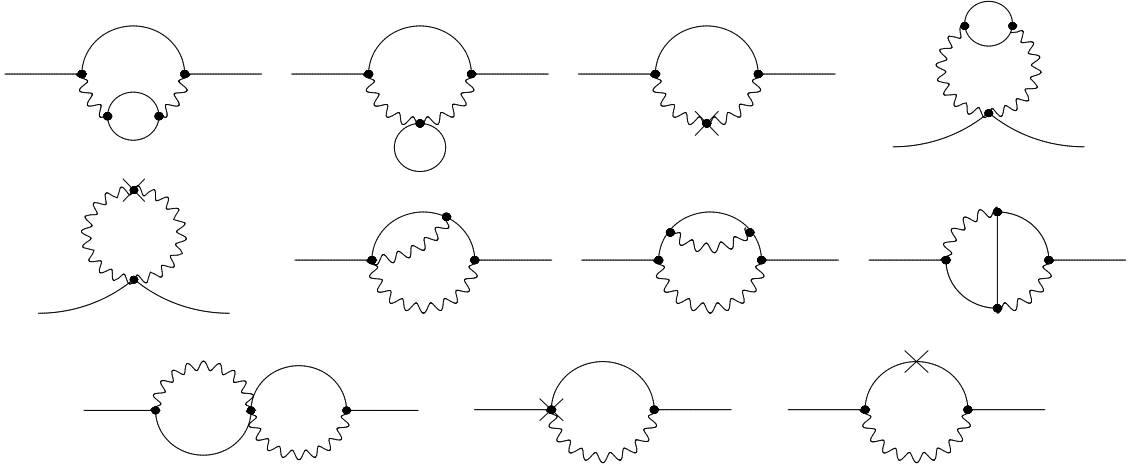}
\caption{Two-loop self-energy graphs for matter superfield.}
\label{Figure_Gamma_Diagrams}
\end{figure}

It is important to note, that there are some graphs, containing an internal
loop or insersions of counterterms on the photon line (first 5 diagrams
in Fig. \ref{Figure_Gamma_Diagrams}), contributing to the two-loop
two-point Green function of the matter superfield. According to \cite{tmf2}
their contribution in Euclidean space is

\begin{eqnarray}\label{Gamma_Oo}
&& \int d^4\theta\,\frac{d^4p}{(2\pi)^4}\,
\Big(\phi^*\phi+\tilde\phi^*\tilde\phi\Big)
\int\frac{d^4k}{(2\pi)^4}
\frac{e^2}{2 k^2 (k+p)^2 \Big(1 + k^{2n}/\Lambda^{2n}\Big)^2}
\times\nonumber\\
&& \times
\Bigg[\Big(1+k^{2n}/\Lambda^{2n}\Big)
\Bigg(1+\frac{e^2}{4\pi^2}\Big(\ln\frac{\Lambda}{\mu}+b_1\Big)\Bigg)
-\nonumber\\
&& - 2 e^2 \Bigg(\int \frac{d^4t}{(2\pi)^4}\,\frac{1}{t^2 (k+t)^2}
- \sum\limits_i c_i \int \frac{d^4t}{(2\pi)^4}\,
\frac{1}{\Big(t^2+M_i^2\Big) \Big((k+t)^2+M_i^2\Big)}
\Bigg)\Bigg].\qquad
\end{eqnarray}

\noindent
and contains only the first degree of $\ln\Lambda$. By comparing
(\ref{Beta_Oo}) and (\ref{Gamma_Oo}) we find that, the limit
(\ref{Beta_Oo}) exists and is equal to $-1/16\pi^2$ multiplied by
the corresponding two-loop contribution to the anomalous dimension. This,
in turn, yields the following contribution to the $\beta$-function:

\begin{equation}
\Delta\beta = -\frac{\alpha^2}{\pi}\Delta\gamma,
\end{equation}

\noindent
where $\Delta\gamma$ is a contribution to the anomalous dimension from
a one-loop diagram and two-loop diagrams, containing corrections to the
photon propagator.

\bigskip

4. In order to calculate the divergent part of $f_O$ we prove,
that the limit

\begin{equation}\label{O_Limit}
\Lambda\frac{d}{d\Lambda} f_O\Big|_{p=0},
\end{equation}

\noindent
where $f_O(\Lambda/p)$ is given by (\ref{O_Diagrams}), exists.
Using the identities

\begin{eqnarray}
&& (k+l)^2 = (q+k+l)^2 + q^2 - (q+k)^2 - (q+l)^2 + k^2 + l^2;\nonumber\\
&& (2q+k+l)^2 = (q+k+l)^2 + (q+k)^2 + (q+l)^2 - k^2 - l^2 + q^2
\end{eqnarray}

\noindent
in Euclidean space it is possible to present (\ref{O_Limit})
as follows:

\begin{eqnarray}\label{O_Limit2}
&& \Lambda\frac{d}{d\Lambda} f_O\Big|_{p=0}
= \Lambda\frac{d}{d\Lambda}\int \frac{d^4k}{(2\pi)^4}
\frac{d^4l}{(2\pi)^4}\frac{d^4q}{(2\pi)^4}\,
\frac{2 e^4}{k^2 \Big(1+k^{2n}/\Lambda^{2n}\Big)\,
l^2\Big(1+l^{2n}/\Lambda^{2n}\Big)}
\times\quad\nonumber\\
&& \times
\frac{1}{q^4 (q+l)^2}
\Bigg\{\frac{2}{(q+k)^2}
-\frac{2 k^2}{(q+k)^4}
- \frac{12 k^2}{(q+k)^2 (q+k+l)^2}
+ \frac{4 (q+l)^2}{(q+k)^2 (q+k+l)^2}
+\nonumber\\
&&
+ \frac{2 (k^4+k^2 l^2)}{(q+k)^4 (q+k+l)^2}
+ \frac{2 k^2 (k+l)^2}{(q+k)^2 (q+k+l)^4}
- \frac{(k+l)^2}{(q+k+l)^4} \Bigg\}.
\end{eqnarray}

\noindent
This expression can be simplified by identities
(\ref{Wonderful_Identity1}) -- (\ref{Wonderful_Identity3}),
presented in appendix \ref{Appendix_Integrals}:

\begin{eqnarray}\label{O_Last_Integral}
&& \Lambda\frac{d}{d\Lambda} f_O\Big|_{p=0}
=\\
&& = \frac{1}{4\pi^2}\Lambda\frac{d}{d\Lambda}
\int \frac{d^4k}{(2\pi)^4} \frac{d^4l}{(2\pi)^4}\,
\frac{e^4}{\Big(1+k^{2n}/\Lambda^{2n}\Big)\,
\Big(1+l^{2n}/\Lambda^{2n}\Big)} \Bigg\{
\frac{1}{2 k^4 l^4} - \frac{1}{k^4 l^2 (k+l)^2}
\Bigg\}.\quad\nonumber
\end{eqnarray}

\noindent
This integral is a finite constant (see appendix \ref{Appendix_Integrals}).
However in order to relate the three-loop $\beta$-function with the two-loop
anomalous dimension, it is convenient to rewrite (\ref{O_Last_Integral})
in the following form:

\begin{eqnarray}\label{FO_Identity}
&&\hspace*{-5mm}
\frac{1}{16\pi^2}\Lambda\frac{d}{d\Lambda}\Bigg(
- \int \frac{d^4k}{(2\pi)^4}\,\frac{d^4l}{(2\pi)^4}\,
\frac{2 e^4}{k^2 l^2 (k+p)^2 (l+p)^2
\Big(1+ k^{2n}/\Lambda^{2n}\Big)
\Big(1+ l^{2n}/\Lambda^{2n}\Big)}
-\nonumber\\
&&\hspace*{-5mm}
- \int \frac{d^4k}{(2\pi)^4}\,\frac{d^4l}{(2\pi)^4}\,
\frac{4 e^4}{\displaystyle k^2 l^2 (l+p)^2 (k+l+p)^2
\Big(1+ k^{2n}/\Lambda^{2n}\Big)
\Big(1+ l^{2n}/\Lambda^{2n}\Big)}
+\nonumber\\
&&\hspace*{-5mm}
+ \int \frac{d^4k}{(2\pi)^4}\,\frac{d^4l}{(2\pi)^4}\,
\frac{4 e^4 (k+l+2p)^2}{k^2 (k+p)^2 l^2 (l+p)^2 (k+l+p)^2
\Big(1+ k^{2n}/\Lambda^{2n}\Big)
\Big(1+ l^{2n}/\Lambda^{2n}\Big)}\Bigg)\Bigg|_{p=0},\nonumber\\
\end{eqnarray}

\noindent
because in this case (\ref{Beta_Oo}) and (\ref{FO_Identity}) give

\begin{eqnarray}
\Lambda \frac{d}{d\Lambda} \Big(f_O+f_{Oo}\Big)\Bigg|_{p=0}
= \frac{1}{16\pi^2} \Lambda \frac{d}{d\Lambda}
\Big(z_1+z_2+\frac{1}{2}z_1^2\Big)\Bigg|_{p=0},
\end{eqnarray}

\noindent
where $z_1$ and $z_2$ are defined by (\ref{Z1_Definition}) and
(\ref{Z2_Definition}) respectively.

\bigskip

5. $f_{Oo_{PV}}$ is finite. Indeed, in Euclidean space

\begin{eqnarray}\label{FOo}
&& f_{Oo_{PV}}= - \sum\limits_j c_j\,
\int \frac{d^4k}{(2\pi)^4}\frac{d^4q}{(2\pi)^4}\,
\frac{e^2}{k^2 \Big(1 + k^{2n}/\Lambda^{2n}\Big)^2}
\times\nonumber\\
&& \times
\Bigg(\frac{(k+p+q)^2+q^2-k^2-p^2}{
\Big((k+q)^2+M_j^2\Big) \Big((k+p+q)^2+M_j^2\Big) \Big(q^2+M_j^2\Big)
\Big((q+p)^2+M_j^2\Big)}
-\nonumber\\
&& \qquad\qquad\qquad\qquad\qquad\qquad
- \frac{4 M_j^2}{\Big((k+q)^2+M_j^2\Big) \Big(q^2+M_j^2\Big)^2
\Big((q+p)^2+M_j^2\Big)} \Bigg)
\times\nonumber\\
&&\times
\Bigg[\Big(1+k^{2n}/\Lambda^{2n}\Big)
\Bigg(1+\frac{e^2}{4\pi^2}\Big(\ln\frac{\Lambda}{\mu}+b_1\Big)\Bigg)
- 2 e^2 \Bigg(\int \frac{d^4t}{(2\pi)^4}\,\frac{1}{t^2 (k+t)^2}
-\nonumber\\
&& - \sum\limits_i c_i \int \frac{d^4t}{(2\pi)^4}\,
\frac{1}{\Big(t^2+M_i^2\Big) \Big((k+t)^2+M_i^2\Big)}
\Bigg)\Bigg].
\end{eqnarray}

\noindent
This expression can be written as

\begin{equation}
f_{Oo_{PV}} = f_1(p/\Lambda) + f_2(p/\Lambda)\,\ln\frac{\Lambda}{\mu},
\end{equation}

\noindent
where the functions $f_1$ and $f_2$ can be easily found from (\ref{FOo}).
In appendix \ref{Appendix_Integrals} we prove, that

\begin{equation}
\int \frac{d^4q}{(2\pi)^4}\,
\Bigg(\frac{(k+q)^2+q^2-k^2}{
\Big((k+q)^2+M^2\Big)^2 \Big(q^2+M^2\Big)^2}
- \frac{4 M^2}{\Big((k+q)^2+M^2\Big) \Big(q^2+M^2\Big)^3}\Bigg) = 0.
\end{equation}

\noindent
and therefore $f_1(0)=f_2(0)=0$. Since the functions $f_1$ and $f_2$
are evidently holomorphic at $p^2=0$, this means that

\begin{equation}
\lim\limits_{\Lambda\to\infty} f_{Oo_{PV}} = 0.
\end{equation}

\bigskip

6. The finiteness of $f_{O_{PV}}$ can be proven similarly. Indeed, it is
evident, that

\begin{equation}
f_{O_{PV}} = f(p/\Lambda).
\end{equation}

\noindent
However, at $p=0$ identities (\ref{PV_Identity1}) -- (\ref{PV_Identity6}),
presented in appendix \ref{Appendix_Integrals}, give $f(0)=0$. Therefore,
$f_{O_{PV}}$ is finite and it vanishes in the limit of the regularization
removed.


\section{Calculation of three-loop integrals, regularized by higher
derivatives.}
\hspace{\parindent}
\label{Appendix_Integrals}

The integral

\begin{equation}
I_{Oo} \equiv \int d^4q\,\frac{q_\mu (k_\mu+q_\mu)}{(k+q)^4 q^4},
\end{equation}

\noindent
can be calculated in four-dimensional spherical coordinates
$(q,\theta_1,\theta_2,\varphi)$. In these coordinates we have

\begin{eqnarray}
&& I_{Oo} = \int d\Omega\,\int\limits_0^\infty dq\,
\frac{q+k\cos\alpha}{\Big(q^2+2qk\cos\alpha+k^2\Big)^2}
=\nonumber\\
&& \qquad\qquad\qquad
= -\frac{1}{2}\int d\Omega\,\int\limits_0^\infty dq\,
\frac{d}{dq} \frac{1}{\Big(q^2+2qk\cos\alpha+k^2\Big)}
= \frac{1}{2}\int d\Omega\,\frac{1}{k^2} = \frac{\pi^2}{k^2},\qquad\quad
\end{eqnarray}

\noindent
where $\alpha$ denotes an angle between four-vectors $k$ and $q$.
This angle can be chosen equal to $\theta_1$, while

\begin{equation}
d\Omega = d\theta_1\,d\theta_2\,d\varphi\,\sin^2\theta_1\sin\theta_2.
\end{equation}

\bigskip

An integral in (\ref{FX2_At_P=0}) and (\ref{FOo}) at $p=0$

\begin{equation}
I_{X2} \equiv
\int\frac{d^4q}{(2\pi)^4}\,
\Bigg\{\frac{2 q_\mu (q_\mu+k_\mu)}{\Big((k+q)^2+M^2\Big)^2
\Big(q^2+M^2\Big)^2} - \frac{4 M^2}{\Big((k+q)^2+M^2\Big)
\Big(q^2+M^2\Big)^3} \Bigg\}.\qquad
\end{equation}

\noindent
can be computed similarly: In the four-dimensional spherical coordinates

\begin{eqnarray}
&& I_{X2} =
\frac{1}{(2\pi)^4}\int\limits_0^\infty dq\,q^3 \int d\Omega\,
\Bigg\{- q\frac{d}{dq} \frac{1}{\Big((k+q)^2+M^2\Big)}
\frac{1}{\Big(q^2+M^2\Big)^2}
-\nonumber\\
&& \qquad\qquad\qquad\qquad\qquad\qquad\qquad\qquad\qquad
- \frac{4 M^2}{\Big((k+q)^2+M^2\Big)
\Big(q^2+M^2\Big)^3} \Bigg\}
=\qquad\nonumber\\
&& = \frac{1}{(2\pi)^4}\int\limits_0^\infty dq \int d\Omega\,
\frac{1}{\Big((k+q)^2+M^2\Big)}
\Bigg\{
\frac{d}{dq} \frac{q^4}{\Big(q^2+M^2\Big)^2}
- \frac{4 M^2 q^3}{\Big(q^2+M^2\Big)^3} \Bigg\} = 0.\qquad\
\end{eqnarray}

\bigskip

In order to compute integral (\ref{O_Last_Integral}) we take into
account its symmetry with respect to the substitution $k\leftrightarrow l$
and present (\ref{O_Last_Integral}) in the following form:

\begin{eqnarray}
&& \Lambda\frac{d}{d\Lambda} f_{O}\Big|_{p=0}
=\nonumber\\
&& = -\frac{1}{4\pi^2}\Lambda\frac{d}{d\Lambda}
\int \frac{d^4k}{(2\pi)^4} \frac{d^4l}{(2\pi)^4}\,
\frac{e^4}{\Big(1+k^{2n}/\Lambda^{2n}\Big)\,
\Big(1+l^{2n}/\Lambda^{2n}\Big)} \frac{k^2+l^2-(k+l)^2}{2 k^4 l^4 (k+l)^2}
=\qquad\nonumber\\
&& = \frac{n}{\pi^2} \int \frac{d^4k}{(2\pi)^4} \frac{d^4l}{(2\pi)^4}\,
\frac{e^4\,l^{2n}/\Lambda^{2n}}{\Big(1+k^{2n}/\Lambda^{2n}\Big)\,
\Big(1+l^{2n}/\Lambda^{2n}\Big)^2}\,\frac{k_\mu l_\mu}{k^4 l^4 (k+l)^2}
\end{eqnarray}

\noindent
The integral over $d^4k$ can be calculated in the four-dimensional
spherical coordinates if the fourth axis is directed along $l_\mu$:

\begin{eqnarray}
&& \Lambda\frac{d}{d\Lambda} f_{O}\Big|_{p=0}
= \frac{n}{\pi^2} \int \frac{d^4l}{(2\pi)^4}\,\frac{1}{(2\pi)^4}
\int\limits_0^\infty dk\,\int\limits_0^\pi d\theta_1\,\sin^2\theta_1
\int\limits_0^\pi d\theta_2\,\sin\theta_2\int\limits_0^{2\pi}d\varphi\,
\times\qquad\nonumber\\
&&\times
\frac{e^4\,l^{2n}/\Lambda^{2n}}{\Big(1+k^{2n}/\Lambda^{2n}\Big)\,
\Big(1+l^{2n}/\Lambda^{2n}\Big)^2}\,
\frac{\cos\theta_1}{l^3 \Big(k^2+2kl\cos\theta_1+l^2\Big)}.
\end{eqnarray}

\begin{figure}[h]

\hspace*{3cm}
\begin{picture}(0,0)(0,0)
\put(7.5,-3.9){$\mbox{Re}(x)$}
\put(4.8,-0.9){$\mbox{Im}(x)$}
\put(3.2,-4.1){$-1$}
\put(5.6,-4.1){$1$}
\put(2,-3.9){$x_0$}
\end{picture}

\hspace*{3cm}
\epsfxsize9.0truecm\epsfbox{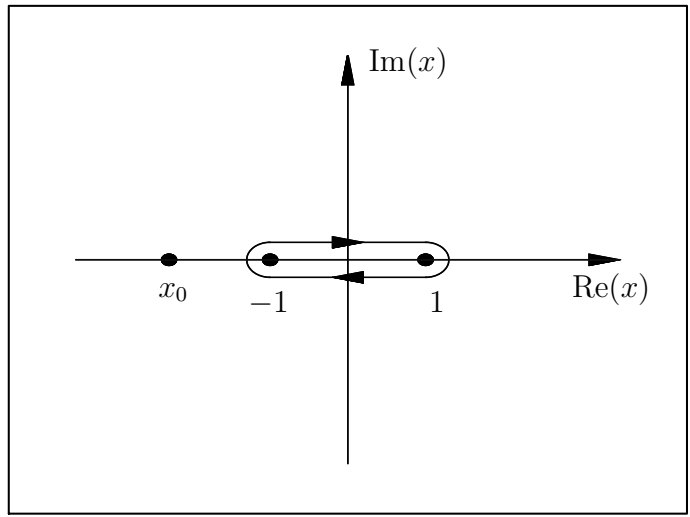}

\caption{Contour $C$ for calculation of the integral over $x$.}
\label{Figure_Contour}
\end{figure}

\noindent
After the substitution $x=\cos\theta_1$ the integral over angles
is reduced to the integral over contour $C$, presented in Fig.
\ref{Figure_Contour}:

\begin{eqnarray}
&& 4\pi\int\limits_{-1}^{1} dx\,\frac{x \sqrt{1-x^2}}{k^2+2klx+l^2}
= 2\pi \oint\limits_C dx\,\frac{x\sqrt{1-x^2}}{k^2+2klx+l^2}
=\nonumber\\
&& = 4\pi^2 i\, \mbox{Res}\Bigg(\frac{x\sqrt{1-x^2}}{k^2+2klx+l^2},
\,x=\infty\Bigg)
-\nonumber\\
&& \qquad\qquad\qquad\qquad\qquad\qquad
- 4\pi^2 i\,\mbox{Res}\Bigg(\frac{x\sqrt{1-x^2}}{k^2+2klx+l^2},
\,x=-\frac{k^2+l^2}{2kl}\Bigg)
=\qquad\nonumber\\
&& = 4\pi^2 i\Bigg(- \frac{i}{4 kl} + \frac{i (k^2+l^2)^2}{8 k^3 l^3}
- \frac{i|k^2-l^2| (k^2 + l^2)}{8 k^3 l^3}\Bigg).
\end{eqnarray}

\noindent
Therefore

\begin{eqnarray}
2\pi \oint\limits_C dx\,\frac{x\sqrt{1-x^2}}{k^2+2klx+l^2} =
\left\{
\begin{array}{l}
{\displaystyle - \frac{\pi^2 l}{k^3},\quad k\ge l;}\\
\\
{\displaystyle - \frac{\pi^2 k}{l^3},\quad l\ge k}
\end{array}
\right.
\end{eqnarray}

\noindent
and

\begin{eqnarray}
&& \Lambda\frac{d}{d\Lambda} f_{O}\Big|_{p=0}
= -\frac{n e^4}{8\pi^2 (2\pi)^4} \int\limits_0^\infty dl\,
\frac{l^{2n}/\Lambda^{2n}}{\Big(1+l^{2n}/\Lambda^{2n}\Big)^2}
\times\nonumber\\
&&\qquad\qquad\qquad\quad\ \ \times
\Bigg(\int\limits_l^\infty dk\,
\frac{l}{k^3 \Big(1+ k^{2n}/\Lambda^{2n}\Big)}
+ \int\limits_0^l dk\,
\frac{k}{l^3 \Big(1+ k^{2n}/\Lambda^{2n}\Big)}
\Bigg).\qquad
\end{eqnarray}

\noindent
Making the substitutions $x = l^2/\Lambda^2$; $y = \Lambda^2/k^2$
in the first integral and $x = \Lambda^2/l^2$; $y = k^2/\Lambda^2$
in the second one, we finally obtain

\begin{eqnarray}
&& \Lambda\frac{d}{d\Lambda} f_{O}\Big|_{p=0}
= - \frac{n e^4}{32\pi^2 (2\pi)^4} \int\limits_0^\infty dx\,
\frac{x^{n}}{(1+x^{n})^2}
\int\limits_0^{1/x}
\frac{dy}{1+ y^{-n}}
-\nonumber\\
&& \qquad\qquad\qquad\qquad\qquad\qquad\qquad
-\frac{n e^4}{32\pi^2 (2\pi)^4} \int\limits_0^\infty dx\,
\frac{x^{n}}{(1+x^{n})^2}
\int\limits_0^{1/x}
\frac{dy}{1+ y^{n}}
=\qquad\nonumber\\
&&
= -\frac{n e^4}{32\pi^2 (2\pi)^4} \int\limits_0^\infty dx\,
\frac{x^{n-1}}{(1+x^{n})^2}
= -\frac{e^4}{32\pi^2 (2\pi)^4} = -\frac{\alpha^2}{32\pi^4}.
\end{eqnarray}

Below we also present the identities, which were used for taking integrals,
which contain higher derivatives.

1. Identities required for calculation of diagrams with
$\phi$- and $\tilde\phi$-lines:

\begin{eqnarray}\label{Wonderful_Identity1}
&& I_1\equiv \int \frac{d^4k}{(2\pi)^4}
\frac{d^4l}{(2\pi)^4}\frac{d^4q}{(2\pi)^4}\,
\frac{(k+q)^2+q^2-k^2}{k^2 \Big(1+k^{2n}/\Lambda^{2n}\Big)\,
l^2\Big(1+l^{2n}/\Lambda^{2n}\Big)\,q^4 (q+k)^4 (q+l)^2}
=\nonumber\\
&& = \frac{1}{16\pi^2}
\int \frac{d^4k}{(2\pi)^4}
\frac{d^4l}{(2\pi)^4}\,
\frac{1}{\Big(1+k^{2n}/\Lambda^{2n}\Big)\,
\Big(1+l^{2n}/\Lambda^{2n}\Big)\,k^4 l^4};\\
&& \vphantom{\Big(}\nonumber\\
\label{Wonderful_Identity2}
&& I_2\equiv \int \frac{d^4k}{(2\pi)^4}
\frac{d^4l}{(2\pi)^4}\,\frac{d^4q}{(2\pi)^4}\,
\frac{1}{k^2 \Big(1+k^{2n}/\Lambda^{2n}\Big)\,
l^2\Big(1+l^{2n}/\Lambda^{2n}\Big) q^4 (q+l)^2}
\times\nonumber\\
&& \times
\Bigg\{\frac{12 k^2}{(q+k)^2 (q+k+l)^2}
- \frac{2 (k^4+k^2 l^2)}{(q+k)^4 (q+k+l)^2}
- \frac{2 k^2 (k+l)^2}{(q+k)^2 (q+k+l)^4} \Bigg\}
=\nonumber\\
&& = \frac{1}{4\pi^2}
\int \frac{d^4k}{(2\pi)^4}
\frac{d^4l}{(2\pi)^4}\,
\frac{1}{\Big(1+k^{2n}/\Lambda^{2n}\Big)\,
\Big(1+l^{2n}/\Lambda^{2n}\Big)\,k^4 l^2 (k+l)^2};\\
&&\vphantom{\Big(} \nonumber\\
\label{Wonderful_Identity3}
&& I_3\equiv \int \frac{d^4k}{(2\pi)^4}
\frac{d^4l}{(2\pi)^4}\frac{d^4q}{(2\pi)^4}
\frac{(k+q+l)^2+q^2-(k+l)^2}{k^2 \Big(1+k^{2n}/\Lambda^{2n}\Big)\,
l^2\Big(1+l^{2n}/\Lambda^{2n}\Big)\,q^4 (q+k+l)^4 (q+k)^2}
=\nonumber\\
&& = \frac{1}{16\pi^2}
\int \frac{d^4k}{(2\pi)^4}
\frac{d^4l}{(2\pi)^4}\,
\frac{1}{\Big(1+k^{2n}/\Lambda^{2n}\Big)\,
\Big(1+l^{2n}/\Lambda^{2n}\Big)}
\Bigg\{\frac{2}{k^4 l^2 (k+l)^2} - \frac{1}{k^4 l^4}  \Bigg\}
\quad\nonumber\\
\end{eqnarray}

2. Identities, required for calculation of diagrams with internal
Pauli-Villars lines:

\begin{eqnarray}\label{PV_Identity1}
&& J_1\equiv \int \frac{d^4k}{(2\pi)^4} \frac{d^4l}{(2\pi)^4}
\frac{d^4q}{(2\pi)^4}
\frac{1}{k^2 \Big(1+k^{2n}/\Lambda^{2n}\Big)
\,l^2 \Big(1+l^{2n}/\Lambda^{2n}\Big) \Big((q+l)^2+M^2\Big)}
\times\nonumber\\
&& \times
\Bigg\{
\frac{1}{(q^2+M^2)^2 \Big((q+k)^2+M^2\Big)}
+ \frac{1}{(q^2+M^2) \Big((q+k)^2+M^2\Big)^2}
-\nonumber\\
&& - \frac{k^2+2M^2}{(q^2+M^2)^2 \Big((q+k)^2+M^2\Big)^2}
- \frac{2M^2}{(q^2+M^2)^3 \Big((q+k)^2+M^2\Big)}
\Bigg\} = 0;\\
&& \vphantom{\Big(}\nonumber\\
\label{PV_Identity2}
&& J_2\equiv \int \frac{d^4k}{(2\pi)^4} \frac{d^4l}{(2\pi)^4}
\frac{d^4q}{(2\pi)^4}
\frac{1}{k^2 \Big(1+k^{2n}/\Lambda^{2n}\Big)
\,l^2 \Big(1+l^{2n}/\Lambda^{2n}\Big) (q^2+M^2)^2}
\times\nonumber\\
&& \times
\frac{1}{\Big((q+k)^2+M^2\Big) \Big((q+l)^2+M^2\Big)}
\Bigg\{
-\frac{k^2 (k^2+l^2)}{\Big((q+k)^2+M^2\Big)
\Big((q+k+l)^2+M^2\Big)}
-\nonumber\\
&&
-\frac{2 M^2 (k^2+l^2)}{\Big((q+k)^2+M^2\Big)
\Big((q+k+l)^2+M^2\Big)}
-\frac{4 M^2 k^2}{(q^2+M^2) \Big((q+k+l)^2+M^2\Big)}
+\nonumber\\
&&
+\frac{6k^2}{\Big((q+k+l)^2+M^2\Big)}
-\frac{k^2 (k+l)^2 +2 M^2 k^2}{\Big((q+k+l)^2+M^2\Big)^2}
\Bigg\}=0;\\
&& \vphantom{\Big(}\nonumber\\
\label{PV_Identity3}
&& J_3\equiv \int \frac{d^4k}{(2\pi)^4} \frac{d^4l}{(2\pi)^4}
\frac{d^4q}{(2\pi)^4}
\frac{1}{k^2 \Big(1+k^{2n}/\Lambda^{2n}\Big)
\,l^2 \Big(1+l^{2n}/\Lambda^{2n}\Big) (q^2+M^2)^2}
\times\nonumber\\
&& \times
\frac{1}{\Big((q+k)^2+M^2\Big) \Big((q+k+l)^2+M^2\Big)}
\Bigg\{
3 - \frac{(k+l)^2+2M^2}{\Big((q+k+l)^2+M^2\Big)}
- \frac{4M^2}{(q^2+M^2)}
+\nonumber\\
&& +\frac{(q+k+l)^2+M^2}{\Big((q+l)^2+M^2\Big)}
-\frac{k^2+2M^2}{\Big((q+k)^2+M^2\Big)}
\Bigg\} = 0;\\
&& \vphantom{\Big(}\nonumber\\
\label{PV_Identity4}
&& J_4\equiv \int \frac{d^4k}{(2\pi)^4} \frac{d^4l}{(2\pi)^4}
\frac{d^4q}{(2\pi)^4}
\frac{1}{k^2 \Big(1+k^{2n}/\Lambda^{2n}\Big)
\,l^2 \Big(1+l^{2n}/\Lambda^{2n}\Big) (q^2+M^2)^2}
\times\nonumber\\
&& \times
\frac{1}{\Big((q+k)^2+M^2\Big)^2 \Big((q+k+l)^2+M^2\Big)}
\Bigg\{
4M^2 - \frac{M^2 (k+l)^2+2M^4}{\Big((q+k+l)^2+M^2\Big)}
+\nonumber\\
&& + \frac{2M^2 (q^2+M^2)}{\Big((q+k)^2+M^2\Big)}
- \frac{2 M^2 k^2+4M^4}{\Big((q+k)^2+M^2\Big)}
- \frac{4M^4}{(q^2+M^2)}
\Bigg\} = 0;\\
&& \vphantom{\Big(}\nonumber\\
\label{PV_Identity5}
&& J_5\equiv \int \frac{d^4k}{(2\pi)^4} \frac{d^4l}{(2\pi)^4}
\frac{d^4q}{(2\pi)^4}
\frac{1}{k^2 \Big(1+k^{2n}/\Lambda^{2n}\Big)
\,l^2 \Big(1+l^{2n}/\Lambda^{2n}\Big) \Big((q+l)^2+M^2\Big)}
\times\nonumber\\
&& \times
\Bigg\{
\frac{2M^2}{\Big(q^2+M^2\Big)^3 \Big((q+k)^2+M^2\Big)}
+\frac{M^2}{\Big(q^2+M^2\Big)^2 \Big((q+k)^2+M^2\Big)^2}
-\nonumber\\
&& -\frac{2M^4+M^2k^2}{\Big(q^2+M^2\Big)^3 \Big((q+k)^2+M^2\Big)^2}
-\frac{3M^4}{\Big(q^2+M^2\Big)^4 \Big((q+k)^2+M^2\Big)}
\Bigg\} = 0;\\
&& \vphantom{\Big(}\nonumber\\
\label{PV_Identity6}
&& J_6\equiv \int \frac{d^4k}{(2\pi)^4} \frac{d^4l}{(2\pi)^4}
\frac{d^4q}{(2\pi)^4}
\frac{1}{k^2 \Big(1+k^{2n}/\Lambda^{2n}\Big)
\,l^2 \Big(1+l^{2n}/\Lambda^{2n}\Big) (q^2+M^2)^2}
\times\nonumber\\
&& \times
\frac{1}{\Big((q+k)^2+M^2\Big) \Big((q+l)^2+M^2\Big)}
\Bigg\{
-\frac{2M^4}{(q^2+M^2)\Big((q+k+l)^2+M^2\Big)}
+\nonumber\\
&&
+\frac{M^2 \Big((q+l)^2+M^2\Big) - M^4}{\Big((q+k+l)^2+M^2\Big)^2}
-\frac{M^2 k^2}{\Big((q+k+l)^2+M^2\Big)^2}
+\frac{2M^2}{\Big((q+k+l)^2+M^2\Big)}
-\nonumber\\
&&
-\frac{M^2 k^2+2M^4}{\Big((q+k)^2+M^2\Big)\Big((q+k+l)^2+M^2\Big)}
\Bigg\} = 0.
\end{eqnarray}

As an example we prove identity (\ref{Wonderful_Identity1}).
For this purpose we consider

\begin{equation}
I_1 =  \int \frac{d^4k}{(2\pi)^4}\,
\frac{d^4l}{(2\pi)^4}\,\frac{d^4q}{(2\pi)^4}\,
\frac{2 (q^2 +q_\mu k_\mu)}{k^2 \Big(1+k^{2n}/\Lambda^{2n}\Big)\,
l^2\Big(1+l^{2n}/\Lambda^{2n}\Big)\,q^4 (q+k)^4 (q+l)^2}
\end{equation}

\noindent
and write the integral over $d^4q$ in four-dimensional spherical
coordinates:

\begin{eqnarray}
&& I_1 =  \int \frac{d^4k}{(2\pi)^4}\,\frac{d^4l}{(2\pi)^4}\,
\frac{1}{k^2 \Big(1+k^{2n}/\Lambda^{2n}\Big)\,
l^2\Big(1+l^{2n}/\Lambda^{2n}\Big)}
\times\nonumber\\
&& \qquad\qquad\qquad
\times
\frac{1}{(2\pi)^4}\int d\Omega\,\int\limits_0^\infty dq\,
\frac{2 (q +q k \cos\alpha)}{\Big(q^2+2qk\cos\alpha+k^2\Big)^2
\Big(q^2+2ql\cos\beta+l^2\Big)}
=\nonumber\\
&& = - \int \frac{d^4k}{(2\pi)^4}\,\frac{d^4l}{(2\pi)^4}\,
\frac{1}{k^2 \Big(1+k^{2n}/\Lambda^{2n}\Big)\,
l^2\Big(1+l^{2n}/\Lambda^{2n}\Big)}
\times\nonumber\\
&& \qquad\qquad\quad\
\times
\frac{1}{(2\pi)^4}\int d\Omega\,\int\limits_0^\infty dq\,
\frac{1}{\Big(q^2+2ql\cos\beta+l^2\Big)}
\frac{\partial}{\partial q}\frac{1}{\Big(q^2+2qk\cos\alpha+k^2\Big)}.
\qquad\
\end{eqnarray}

\noindent
Performing integrating by parts in the last integral, we obtain

\begin{eqnarray}
&& I_1 = - \int \frac{d^4k}{(2\pi)^4}\,\frac{d^4l}{(2\pi)^4}\,
\frac{1}{k^2 \Big(1+k^{2n}/\Lambda^{2n}\Big)\,
l^2\Big(1+l^{2n}/\Lambda^{2n}\Big)}
\times\nonumber\\
&&\qquad
\times
\frac{1}{(2\pi)^4}\int d\Omega\,\Bigg\{
\frac{1}{\Big(q^2+2qk\cos\alpha+k^2\Big)
\Big(q^2+2ql\cos\beta+l^2\Big)}\Bigg|_0^\infty
-\nonumber\\
&& \qquad\qquad\qquad\qquad
- \int\limits_0^\infty dq\,
\frac{1}{\Big(q^2+2qk\cos\alpha+k^2\Big)}
\frac{\partial}{\partial q}
\frac{1}{\Big(q^2+2ql\cos\beta+l^2\Big)}\Bigg\}
=\nonumber\\
&& = \int \frac{d^4k}{(2\pi)^4}\,\frac{d^4l}{(2\pi)^4}\,
\frac{1}{k^2 \Big(1+k^{2n}/\Lambda^{2n}\Big)\,
l^2\Big(1+l^{2n}/\Lambda^{2n}\Big)}
\times\nonumber\\
&& \ \times
\frac{1}{(2\pi)^4}\int d\Omega\,\Bigg\{
\frac{1}{k^2\,l^2}
- \int\limits_0^\infty dq\,
\frac{2(q+k\cos\beta)}{\Big(q^2+2qk\cos\alpha+k^2\Big)
\Big(q^2+2ql\cos\beta+l^2\Big)^2}\Bigg\}
=\nonumber\\
&& = \int \frac{d^4k}{(2\pi)^4}\,\frac{d^4l}{(2\pi)^4}\,
\frac{1}{k^2 \Big(1+k^{2n}/\Lambda^{2n}\Big)\,
l^2\Big(1+l^{2n}/\Lambda^{2n}\Big)}
\times\nonumber\\
&& \qquad\qquad\qquad\qquad\qquad\qquad\quad
\times
\Bigg\{\frac{1}{8\pi^2}\,\frac{1}{k^2\,l^2}
- \int\frac{d^4q}{(2\pi)^4}
\frac{2(q^2 + q_\mu l_\mu)}{q^4 (q+k)^2 (q+l)^4}\Bigg\}\qquad
\end{eqnarray}

\noindent
The last integral in this expression is evidently equal to $I_1$.
Therefore,

\begin{equation}
I_1 = \frac{1}{16\pi^2}
\int \frac{d^4k}{(2\pi)^4}\,
\frac{d^4l}{(2\pi)^4}\,
\frac{1}{\Big(1+k^{2n}/\Lambda^{2n}\Big)\,
\Big(1+l^{2n}/\Lambda^{2n}\Big)\,k^4 l^4}.
\end{equation}

The other identities can be derived in the similar way.


\pagebreak


\end{document}